\documentclass{sig-alternate-05-2015}

\usepackage{algorithm}
\usepackage{algorithmic}
\usepackage{amsmath}
\usepackage{balance}
\usepackage{cite}
\usepackage{enumitem}
\usepackage{epstopdf}
\usepackage{graphicx}
\usepackage[caption=false]{subfig}
\usepackage{xcolor}
\usepackage{url}

\DeclareMathOperator{\argmax}{argmax}

\makeatletter
\def\@copyrightspace{\relax}
\makeatother

\begin{document}
    
\title{Real-Time Influence Maximization on Dynamic Social Streams}
    
\numberofauthors{1}
\author{
    \alignauthor Yanhao Wang, Qi Fan, Yuchen Li, Kian-Lee Tan\\
    \affaddr{School of Computing, National University of Singapore, Singapore}\\
    \email{\{yanhao90, fanqi, liyuchen, tankl\}@comp.nus.edu.sg}
}
\maketitle
    
\newdef{example}{Example}
\newdef{problem}{Problem}
\newtheorem{definition}{Definition}
\newtheorem{lemma}{Lemma}
\newtheorem{theorem}{Theorem}

\begin{abstract}
    Influence maximization (IM), which selects a set of $k$ users (called seeds)
    to maximize the influence spread over a social network,
    is a fundamental problem in a wide range of applications
    such as viral marketing and network monitoring.
    Existing IM solutions fail to consider the highly dynamic nature of social influence,
    which results in either poor seed qualities or long processing time when the network evolves.
    To address this problem, we define a novel IM query named Stream Influence Maximization (SIM)
    on social streams. Technically, SIM adopts the sliding window model
    and maintains a set of $k$ seeds with the largest influence value
    over the most recent social actions.
    Next, we propose the Influential Checkpoints (IC) framework
    to facilitate continuous SIM query processing.
    The IC framework creates a checkpoint for
    each window slide and ensures an $\varepsilon$-approximate solution.
    To improve its efficiency, we further devise a Sparse Influential Checkpoints (SIC) framework
    which selectively keeps $O(\frac{\log{N}}{\beta})$ checkpoints for a sliding window of size $N$
    and maintains an $\frac{\varepsilon(1-\beta)}{2}$-approximate solution.
    Experimental results on both real-world and synthetic datasets confirm
    the effectiveness and efficiency of our proposed frameworks
    against the state-of-the-art IM approaches.
\end{abstract}

\section{Introduction}
\label{sec:introduction}

Social media advertising has become an indispensable tool
for many companies to promote their business online. 
Such trends have generated over 17 billion dollars advertising revenue for Facebook in 2015
alone\footnote{\url{https://www.bamsec.com/companies/1326801/facebook-inc}}.
\emph{Influence Maximization} (IM) is a key algorithmic problem
behind social media viral marketing~\cite{Domingos2001Mining,Kempe2003Maximizing}.
Through the word-of-mouth propagation among friends,
IM aims to select a set of $k$ users such that the source information (e.g., advertisement)
is maximally spread in the network,
and it has been extensively researched~\cite{Aggarwal2012Influential,Borgs2014MSI,Chen2015OTI,Chen2015Influential,Goyal2011Data,Guo2016Influence,Li2015Conformity,Li2014ELI,Li2015Real,Li2017Discovering,Kempe2003Maximizing,Nguyen2016Stop,Ohsaka2016Dynamic,Tang2014Influence,Tang2015Influence,Wang2016Distance,Zhuang2013Influence} in the last decade.
Besides viral marketing, IM is also the cornerstone in many other important applications
such as network monitoring~\cite{Leskovec2007Cost} and recommendation~\cite{Song2006PRD}.

Most existing IM solutions rely on influence probabilities between users to acquire the seeds.
The influence probabilities are usually derived from
social actions in online social networks~\cite{Goyal2010Learning,Kutzkov2013STRIP,Saito2008Prediction},
e.g., ``retweets'' on Twitter.
In reality, social influences are highly dynamic
and the propagation tendencies between users can be altered drastically
by breaking news and trending topics.
Consequently, the seeds selected by IM methods~\cite{Borgs2014MSI,Kempe2003Maximizing,Nguyen2016Stop,Tang2014Influence,Tang2015Influence}
that assume static social influences can quickly become outdated.
Recently, there are some research efforts on IM under dynamic social influences.
However, existing solutions for dynamic IM
either do not offer theoretical guarantees for the qualities of the seeds selected
(e.g.,~\cite{Aggarwal2012Influential,Zhuang2013Influence})
or provide guarantees at the expense of high processing overhead for updates
(e.g.,~\cite{Chen2015Influential,Ohsaka2016Dynamic}).
In fact, the state-of-the-art dynamic IM solution~\cite{Ohsaka2016Dynamic}
is only able to process several hundred updates per second,
which is far lower than the update rates of real-world social networks.
For example, over 7.4K tweets are generated on Twitter in one second\footnote{\url{http://www.internetlivestats.com/one-second/}}.

To resolve the aforementioned drawbacks and make dynamic IM both effective and efficient,
we propose a novel \emph{Stream Influence Maximization} (SIM) query to
support real-time influential users tracking.
SIM utilizes the widely available social actions
(e.g., ``retweets'' on Twitter, ``replies'' on Reddit and ``comments'' on Facebook)
to estimate the social influence and maintain the seed set continuously.
To capture the short-term memory effect of social influences~\cite{Soda2004Network},
where past influences quickly fade as new influences emerge,
SIM adopts the \emph{sliding window} model~\cite{Datar2002Maintaining}
which always considers the most recent $N$ actions
and strives to find $k$ users who collectively have the largest influence value in the current window.
In addition, SIM supports general monotone submodular functions
to compute the influence values as such functions
are often used to represent the ``diminishing returns'' property of social influences
in different types of IM problems~\cite{Barbieri2012Topic,Chen2015OTI,Goyal2011Data,Kempe2003Maximizing,Li2015Conformity,Li2014ELI,Wang2016Distance}.

Due to the NP-hardness of SIM, we focus on processing it approximately with theoretical bounds.
Leveraging the monotonicity and submodularity of influence functions,
a na{\"i}ve greedy algorithm~\cite{Nemhauser1978Analysis} can provide a $(1-1/e)$ approximate
solution for SIM.
However, the greedy algorithm requires $O(k \cdot |U|)$
($|U|$ is the number of users in the network)
influence function evaluations for each update.
Empirically, it takes around 10 seconds
to select $100$ seeds from a network with $500,000$ users,
which hardly matches the rates of real-world social streams.
Another closely related technique to SIM is
\emph{Streaming Submodular Optimization} (SSO)~\cite{Badanidiyuru2014Streaming,Kumar2015Fast}.
Existing SSO approaches~\cite{Badanidiyuru2014Streaming,Kumar2015Fast} can
provide solutions with theoretical guarantees
for maximizing submodular functions with cardinality constraints over append-only streams.
However, to the best of our knowledge, none of the proposed SSO algorithms
can support the sliding window model.

In this paper, we propose a novel \emph{Influential Checkpoints} (IC) 
framework to support efficient continuous SIM processing with theoretical guarantees.
IC not only tracks the solution for the current window
but also maintains partial solutions called \emph{Checkpoints}
for future windows that overlap with the current window. 
Therefore, for every subsequent window shift, the up-to-date solution can
be retrieved efficiently. 
We further design a generic Set-Stream Mapping (SSM) interface which
can adapt a wide range of existing SSO 
algorithms to SIM so that the solution retrieved for each sliding window has
at least the same approximation ratio as those algorithms.
However, maintaining all $O(N)$ checkpoints incurs significant update overhead
($N$ is the number of actions in a window).
To support efficient IC maintenance,
we propose a \emph{Sparse Influential Checkpoints} (SIC) framework to selectively maintain
a subset of checkpoints by leveraging the monotonicity and subadditivity
of the influence values returned by different checkpoints.
Consequently, SIC only keeps $O(\frac{\log N}{\beta})$ checkpoints
and maintains an $\frac{\varepsilon(1-\beta)}{2}$-approximate solution. 

We hereby summarize our contributions as follows.
\begin{itemize}
    \item We address the limitations of existing IM solutions
    in supporting fast evolving social networks
    and propose a novel SIM query over sliding windows. (Section~\ref{sec:problem})
    \item We develop a novel \emph{Influential Checkpoints} (IC) framework
    for SIM query processing.
    It is integrated with a generic Set-Stream Mapping (SSM) interface
    to incorporate existing $\varepsilon$-approximate SSO algorithms
    while retaining their approximation ratios. (Section~\ref{sec:ic})
    \item We further propose the SIC framework to selectively
    maintain $O(\frac{\log{N}}{\beta})$ checkpoints for a sliding window of size $N$.
    Leveraging the subadditivity and submodularity of the influence values returned
    by different checkpoints,
    an $\frac{\varepsilon(1-\beta)}{2}$-approximation ratio is always guaranteed.
    (Section~\ref{sec:sic})
    \item We experimentally evaluate the effectiveness and efficiency of our proposed frameworks.
    First, the qualities of the seeds selected by IC and SIC are competitive
    with the state-of-the-art IM algorithms on both static and dynamic graphs.
    Second, SIC achieves speedups of up to 2 orders of magnitude over the static approaches.
    Third, SIC achieves up to 8 times speedups over IC
    with less than $5\%$ quality losses. (Section~\ref{sec:experiments})
\end{itemize}

\section{Related work}
\label{sec:relatedwork}
We summarize the most relevant literature from three
areas: influence maximization, streaming submodular optimization and function estimation on sliding windows.
\subsection{Influence Maximization (IM)}
\label{subsec:im}
IM aims to extract a given number of users that maximize the influence spread over a network.
Previous efforts on IM can be generally categorized into static methods and dynamic methods based
on their abilities to handle changes in social influences. Here, we summarize them separately.

\noindent
\textbf{IM in Static Networks}:
There has been a vast amount of literature on influence maximization (IM) in static networks
over the last decade (see~\cite{Borgs2014MSI,Chen2015OTI,Li2015Conformity,Li2014ELI,Li2015Real,Kempe2003Maximizing,Nguyen2016Stop,Tang2014Influence,Tang2015Influence,Wang2016Distance}).
The state-of-the-art static IM method on the classic influence models (i.e., independent cascade (IC) and linear threshold (LT)) is IMM~\cite{Tang2015Influence}.
It runs in nearly linear time wrt.~the graph size with a $(1-1/e-\varepsilon)$ approximation guarantee.
Nevertheless, these static IM methods including IMM
cannot efficiently support highly evolving networks since
a complete rerun is required for every update on influence graphs.

There are also many static methods considering different types of IM problems by
extending classic influence models.
For example, topic-aware IM~\cite{Barbieri2012Topic,Chen2015OTI} considers
the influence diffusion under topic models;
location-aware IM~\cite{Li2014ELI,Wang2016Distance} focuses on maximizing
the influence spread in certain spatial areas;
and conformity-aware IM~\cite{Li2015Conformity} considers users' conformity tendencies
in the influence estimation.

\noindent
\textbf{IM in Dynamic Networks}:
Recently, there are emerging studies about IM in dynamic networks.
However, most of these methods cannot provide a theoretical guarantee of their seed quality and may return arbitrarily bad solutions~\cite{Aggarwal2012Influential,Zhuang2013Influence}.
Chen et al.~\cite{Chen2015Influential} proposed an Upper Bound Interchange (UBI) method
with a $1/2$-approximation ratio. Nevertheless, UBI is sensitive to the number of users selected.
When the size of the seed set increases, both its performance and solution quality degrade dramatically.
This prevents UBI from being practically useful.
Very recently, a new dynamic IM method with a theoretical bound is presented
in~\cite{Ohsaka2016Dynamic}.
It dynamically maintains a RIS-based~\cite{Borgs2014MSI} index against changes on graphs
and achieves a $(1-1/e-\varepsilon)$ approximation guarantee.
However, due to the high maintenance cost, it can only
process several hundred of influence graph updates per second.
This is far from the requirement of real-world social streams.
Therefore, existing dynamic IM methods cannot provide high-quality solutions efficiently.

\subsection{Streaming Submodular Optimization}
Another closely related field to SIM is the Streaming Submodular Optimization (SSO)~\cite{Saha2009Maximum,Ausiello2012Online,Badanidiyuru2014Streaming,Kumar2015Fast,Yu2013SetCoverage}.
SSO adopts the append-only streaming model where
elements arrive one by one and the objective
is to dynamically maintain a set of at most $k$ elements to maximize
a submodular function wrt.~all the observed elements at any time.
Saha et al.~\cite{Saha2009Maximum} and Ausiello et al.~\cite{Ausiello2012Online}
developed two swapping-based approaches for a special case of SSO (i.e., the online Maximum $k$-Coverage problem)
with the same $1/4$ approximation ratio.
Then, Yu and Yuan~\cite{Yu2013SetCoverage} proposed a greedy-based algorithm for the online Maximum $k$-Coverage problem
with an approximation ratio around 0.3 based on computer-aided simulation.
The state-of-the-art SSO solutions are
\textsc{SieveStreaming}~\cite{Badanidiyuru2014Streaming} and \textsc{ThresholdStream}~\cite{Kumar2015Fast}.
They maintain a sequence of candidate solutions by estimating the optimal value
and each solution uses its own threshold to decide whether a new element is included.
At any time, the candidate solution with the largest function among all maintained solutions
is returned as the result.
Both of them can achieve a $(1/2-\beta)$ approximation ratio for any $\beta>0$.

Unfortunately, SSO algorithms cannot be directly applied to the sliding window model
because they do not handle the continuous expiry of elements.
Nevertheless, we will show in Section~\ref{subsec:oracle} that existing
SSO algorithms can serve as checkpoint oracles in the \emph{IC} and \emph{SIC} frameworks.

\subsection{Function Estimation on Sliding Windows}
Several works~\cite{Datar2002Maintaining,Braverman2007Smooth} studied how to continuously
estimate a function in the sliding window model.
They leverage special properties of target functions
to achieve sublinear performance and reasonable quality.
Let $g$ be the target function, and $A$,$B$,$C$ be three sequences on streams
such that $B$ is a tail subsequence of $A$ and $C$ is contiguous to $B$.
The exponential histogram~\cite{Datar2002Maintaining} is proposed to approximate
{\em weakly additive} functions,
i.e., $g(A)+g(C)\leq g(A \cup C) \leq c(g(A)+g(C))$ for some small constant $c$.
The smooth histogram~\cite{Braverman2007Smooth} requires that
the target functions are {\em $(\alpha,\beta)$-smooth}.
Specifically, we say $g$ is {\em $(\alpha,\beta)$-smooth}
if $\frac{g(B)}{g(A)}\geq 1-\beta$,
then $\frac{g(B\cup C)}{g(A\cup C)}\geq 1-\alpha$ for some $0<\beta\leq\alpha<1$.
Following the analysis in~\cite{Braverman2007Smooth},
smooth histograms are applicable only when $g$ can be computed with
an approximation ratio of at least $0.8$ in the append-only streaming setting.
In this paper, we use monotone submodular influence functions~\cite{Kempe2003Maximizing}
widely adopted in the social influence analysis.
However, monotone submodular functions are not {\em weakly additive} and
existing SSO algorithms~\cite{Badanidiyuru2014Streaming,Kumar2015Fast}
can only achieves an approximation of at most $1/2$.
This imply that these techniques cannot be directly applied to our scenario.

\section{Problem Statement}
\label{sec:problem}
We consider a social stream over a social network with
a user set $U$. The social stream comprises
unbounded time-sequenced social actions which are
generated by user activities in the social network.
Let $a_t=\langle u, a_{t'} \rangle_t$ $(t'<t)$ be an action at time $t$
representing the following social activity:
\emph{user $u$ performs $a_t$ at time $t$ responding to an earlier action $a_{t'}$}.
Typical actions include ``retweet'' on Twitter, ``reply'' on Reddit, ``comment'' on Facebook,
to name just a few.
If an action $a_t$ does not respond to any previous action,
e.g., a user $u$ posted an original tweet,
we call it a \emph{root action} and denote it by $a_t=\langle u,nil\rangle_t$.

Like many data streams, social streams are time-sensitive: 
recent actions are more valuable than those in the past.
We adopt the well-recognized \emph{sequence-based sliding window}~\cite{Datar2002Maintaining}
model to capture such essence.
Let $N$ be the window size, a sequence-based sliding window $W_t$ maintains
the latest $N$ actions till $a_t$ in the stream, i.e., $W_t=\{a_{t-N+1},\ldots,a_t\}$.
For simplicity, we use $W_t[i]$ to represent the $i$-th ($i \geq 1$) action within $W_t$.
Then, we use $A_t \subseteq U$ to denote the set of active users
who perform at least one action in $W_t$,
i.e., $A_t = \{W_t[i].u|i=1,\ldots,N\}$.

Since social actions directly reflect the information diffusion
in the social network~\cite{Goyal2010Learning,Goyal2011Data,Subbian2016QTI,Kutzkov2013STRIP},
we define the \emph{influence} between users according to their performed actions.
We say user $u$ \emph{influences} user $v$ in $W_t$, denoted by  $(u \leadsto v)_t$,
if there exists an action $a$ performed by user $v$
s.t. $a \in W_t$ and $a$ is {\it directly} or {\it indirectly} triggered by an action $a'$ of $u$.
It is notable that such an $a'$ is not necessarily in $W_t$.

We formally define the influence set of a user as follows:
\begin{definition}
\label{def:influence}
The influence set of a user $u \in U$ at time $t$,
denoted as $I_t(u) \subseteq A_t$, is the set of users who
are influenced by $u$ wrt.~the sliding window at time $t$ (i.e., $W_t$).
Equivalently, $I_t(u)=\{v|(u \leadsto v)_t\}$.
\end{definition}

Intuitively, 
the influence set of $u$ denotes the set of users
who recently performed actions under the impact of $u$. 
The concept of the influence set can be naturally extended to a set of users.
In particular,
let $S=\{u_1,\ldots,u_k\}$ be a set of $k$ users, the influence set of $S$ wrt.~$W_t$
is a union of the influence sets of all its members, i.e., $I_t(S)=\cup_{u \in S}I_t(u)$.
Then the influence value of $S$ is measured by $f(I_t(S)):2^{|U|} \rightarrow \mathbb{R}_{\geq 0}$.
We consider $f(I_t(\cdot))$ as a nonnegative monotone\footnote{A set function $g$ is monotone if for all $A\subseteq B$, $g(A)\leq g(B)$.}
submodular\footnote{A set function $g$ is submodular if for all $A\subseteq B$, and any element $x \not\in B$, $g(A\cup\{x\})-g(A) \geq g(B\cup\{x\})-g(B)$.}
function~\cite{Nemhauser1978Analysis}
which is widely adopted by many IM problems
for its natural representation of the
``diminishing returns'' property on the social influence~\cite{Kempe2003Maximizing}.

For ease of presentation, we only consider
the cardinality function, i.e., $f(I_t(\cdot))=|I_t(\cdot)|$,
as the influence function in the main text of this paper.
It should be noted that any other monotone submodular influence functions
can also be used in our frameworks.
In Appendix~\ref{sec:discussion}, we will discuss how to extend our frameworks
to different types of IM problems by filtering the streams or using different influence functions.

Example~\ref{exm:def} illustrates our definition of \emph{influence}
over the sliding windows on a social action stream. 
\begin{figure}[t]
    \centering
    \includegraphics[width=0.475\textwidth]{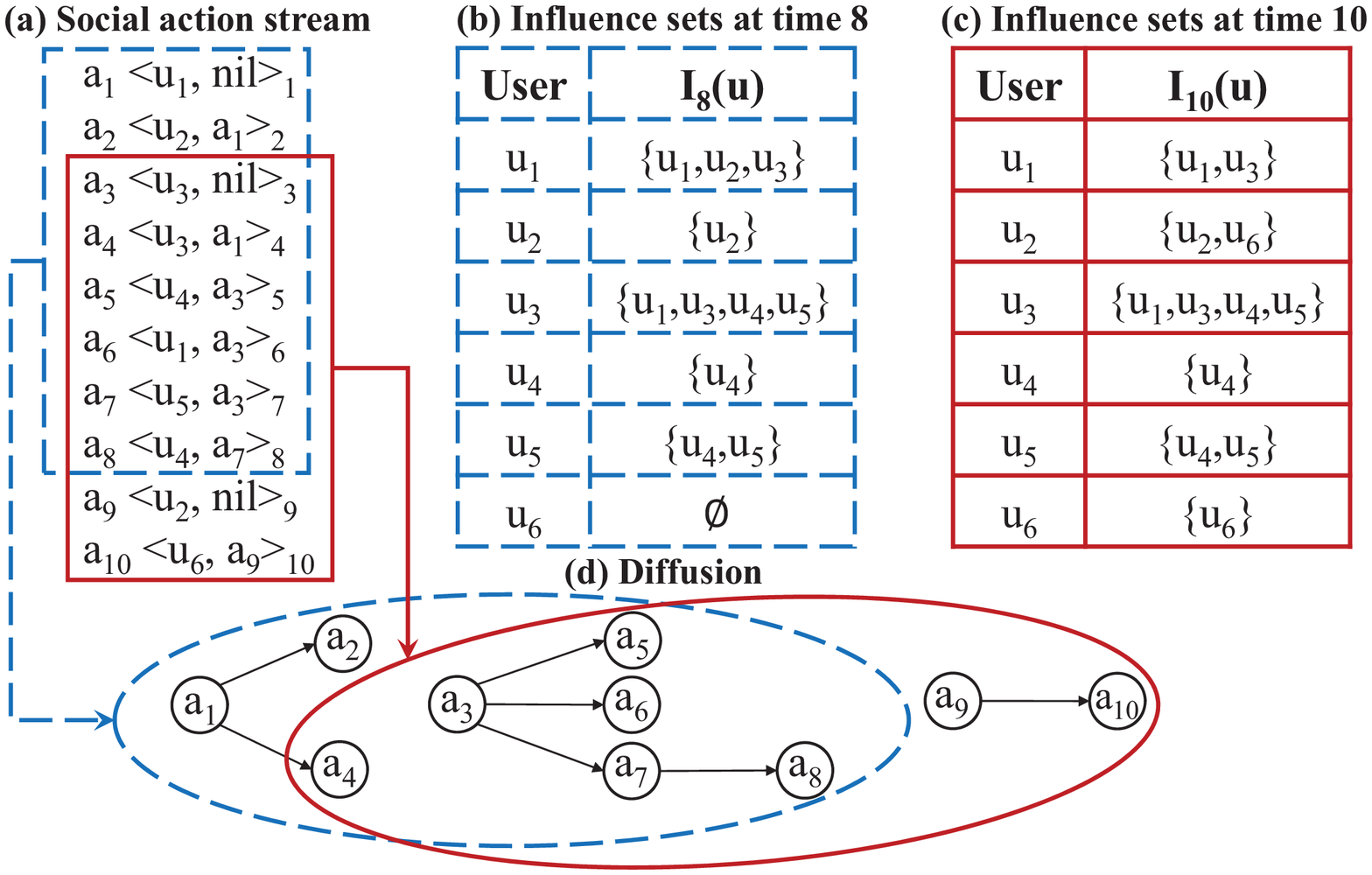}
    \caption{A social action stream and the influences of users over sliding windows.}
    \label{fig:example}
\end{figure}
\begin{example}
    \label{exm:def}
    Figure~\ref{fig:example}(a) gives an example of a social action stream.
    Given the window size $N=8$, two windows, $W_8$ and $W_{10}$,
    are highlighted in blue and red boxes respectively.
    In Figure~\ref{fig:example}(b), $I_8(u_1)=\{u_1,u_2,u_3\}$ as $a_1$,$a_6$ are performed by $u_1$
    and $a_2$,$a_4$ performed by $u_2$,$u_3$ respectively are triggered by $a_1$ in $W_8$.
    When the window slides from $W_8$ to $W_{10}$, $a_1$,$a_2$ expire while $a_9$,$a_{10}$ arrive.
    Then, $I_{10}(u_1)=\{u_1,u_3\}$ as Figure~\ref{fig:example}(c).
    Due to the expiry of $a_2$, $u_2$ is deleted from $I_{10}(u_1)$.
    However, since $a_4$ has not expired yet, $u_1$ still \emph{influences} $u_3$ in $W_{10}$
    regardless of the expiry of $a_1$.
\end{example}

As new actions arrive at high speed while old ones expire at the same rate, 
users with the largest influence values keep evolving.
To track the influential users over social streams in real-time,
we propose a \textbf{\underline{S}}tream \textbf{\underline{I}}nfluence \textbf{\underline{M}}aximization (SIM)
query which is formally defined as follows:
\begin{definition}
Let $W_t$ be the sliding window at time $t$,
Stream Influence Maximization (SIM) is a \emph{continuous} query on a social stream
that returns a set of at most $k$ users $S^{opt}_t$ who collectively achieve
the largest influence value wrt.~$W_t$:
    \begin{equation}
        S^{opt}_t = \argmax_{S \in U \wedge |S| \leq k}{f(I_t(S))}
    \end{equation}
\end{definition}
We continue with the running example in Figure~\ref{fig:example} to show how SIM keeps
track of the most influential users over the sliding windows.
\begin{example}
Given $k=2$, SIM returns $S^{opt}_{8}=\{u_1,u_3\}$ as the most influential users at time $8$
since $I_{8}(S^{opt}_{8}) = I_{8}(u_1) \cup I_{8}(u_3)$ contains all users in $A_8$.
We have $f(I_{8}(S^{opt}_{8}))=5$ using the cardinality function. 
However, as $a_1$,$a_2$ expire while $a_9$,$a_{10}$ arrive at time $10$,
$f(I_{10}(S^{opt}_{8}))=4$ as $u_2$ is deleted from $I_{10}(S^{opt}_{8})$.
Thus, SIM returns $S^{opt}_{10}=\{u_2,u_3\}$ in $W_{10}$
because $I_{10}(S^{opt}_{10})$ contains all users in $A_{10}$.
We have $f(I_{10}(S^{opt}_{10}))=6$ accordingly.
\end{example}

Note that the solutions proposed in this paper also support the case
where the sliding window shifts for more than one action.
For simplicity, we focus on presenting solutions for handling sliding windows
with single action shift at a time
and leave the discussion on handling multiple action shifts at a time
to Section~\ref{subsec:multiple}.

Then, we prove the NP-hardness of SIM by showing that a well-known NP-hard problem,
i.e.,\emph{the Maximum $k$-Coverage problem}~\cite{Ausiello2012Online,Feige1998Threshold,Saha2009Maximum},
is polynomial-time reducible to SIM.
\begin{theorem}
\label{thm:nphard}
    SIM is NP-hard.
\end{theorem}
\begin{proof}
We prove the theorem by reducing a well-known NP-hard problem,
i.e., \emph{Maximum $k$-Coverage}~\cite{Feige1998Threshold}, to SIM.
A \emph{Maximum $k$-Coverage} instance consists of an integer $k$
and a collection of $m$ non-empty sets $\mathcal{S}=\{s_1,s_2,\ldots,s_m\}$.
It aims to find a subset $\mathcal{S'} \subseteq \mathcal{S}$
such that $|\mathcal{S'}|\leq k$ and $|\cup_{s_i \in \mathcal{S'}}s_i|$ is maximized.
Any \emph{Maximum $k$-Coverage} instance is reduced to a \emph{SIM} instance as follows:
Let $t$ be the timestamp and $G=\cup_{s_i \in \mathcal{S}}s_i$ be the ground set.
For any set $s_i \in \mathcal{S}$, we create an action $a_{t'} = \langle u_i, nil \rangle_{t'}$ where $u_i \notin G$.
For each element $e \in s_i$, we create an action $a_{t} = \langle e, a_{t'} \rangle_{t}$($t'<t$).
Next, we stream all actions created to SIM in the ascending order of $t$.
Since the total number of actions created is $O(m \cdot |G|)$,
the reduction is performed in polynomial time.
Let the window size $N$ equal to the total number of actions created.
Let $S^{opt}_N=\{u_i,\cdots,u_j\}$ be the optimal solution of SIM wrt.~$W_N$ using the cardinality function.
$\{s_i,\cdots,s_j\}$ will be optimal for the corresponding \emph{Maximum $k$-Coverage} instance as well.
Thus, the reduction naturally follows.
\end{proof}
\begin{table}[t]
    \centering
    \caption{Frequently used notations}
    \label{tbl:symbol}
    \begin{tabular}{|l|p{155pt}|}
        \hline
        \textbf{Symbol} & \textbf{Definition and Description}\\\hline
        $U$ & the set of all users in a social network\\\hline
        $a_t=\langle u, a_{t'}\rangle_t$ & a user $u$ performs an action at time $t$
        triggered by an action $a_{t'}$($t'<t$)\\\hline
        $N$ & the size of the sliding window\\\hline
        $L$ & the number of actions for each window slide, $L=1$ by default\\\hline
        $W_t,W_t[i]$ & the sliding window at time $t$,
        and the $i$-th action in the window\\\hline
        $I_t(u),I_t(S)$ & the influence set of a user $u$ or a set of users $S$ wrt.~$W_t$\\\hline
        $I_t[i](u),I_t[i](S)$ & the influence set of $u$ or $S$
        for contiguous actions $\{W_t[i],\ldots,W_t[N]\}$\\\hline
        $f$ & a monotone submodular influence function\\\hline
        $k$ & the cardinality constraint of SIM\\\hline
        $\Lambda_t[i]$ & an influential checkpoint maintaining an $\varepsilon$-approximate solution of SIM for $\{W_t[i],\ldots,W_t[N]\}$\\\hline
        $S^{opt}_t,\mathsf{OPT}_t$ & the optimal seed set of SIM wrt.~$W_t$,
        and its influence value $\mathsf{OPT}_t=f(I_t(S^{opt}_t))$\\\hline
        $S^{opt}_t[i],\mathsf{OPT}_t[i]$ & the optimal seed set of SIM for
        $\{W_t[i],$ $\ldots,$ $W_t[N]\}$, and its influence value\\\hline
    \end{tabular}
\end{table}

Before moving on to the technical parts of this paper,
we summarize the frequently used notations in Table~\ref{tbl:symbol}.

\section{Influential Checkpoints}
\label{sec:ic}
\begin{figure*}[ht]
    \centering
    \includegraphics[width=\textwidth]{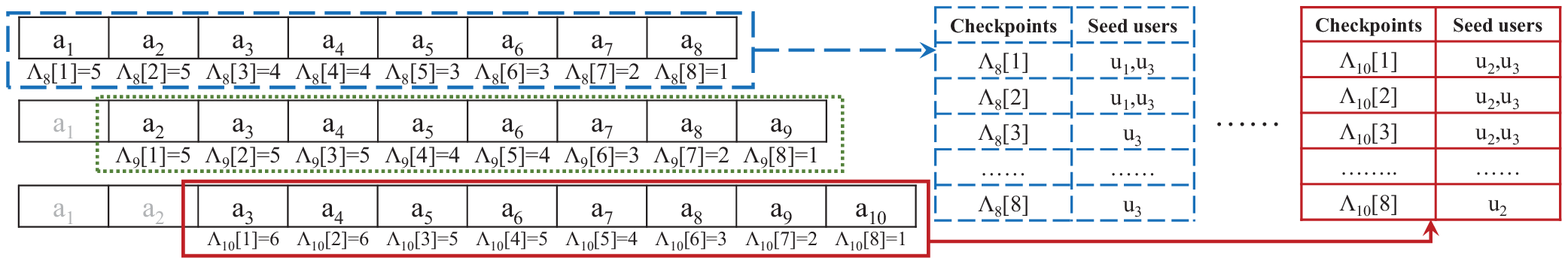}
    \caption{An example of checkpoint maintenance in IC.}
    \label{fig:ic}
\end{figure*}
Since SIM is NP-hard, it is infeasible to maintain the optimal
seed set for each sliding window in polynomial time.
Therefore, our goal is to maintain an approximate solution achieving 
a bounded ratio to the optimal one efficiently. 
A na{\"i}ve scheme is to run the \emph{greedy} algorithm~\cite{Nemhauser1978Analysis} for each window slide.
The \emph{greedy} algorithm starts with an empty user set $S_0=\emptyset$,
and at each iteration $i$ ($1\leq i \leq k$),
it incrementally adds a user $u$ to the partial user set $S_{i-1}$ maximizing
$f(I(S_{i-1}\cup \{u\}))-f(I(S_{i-1}))$ wrt.~$W_t$.
Although the \emph{greedy} algorithm obtains a $(1-1/e)$-approximation ratio
which is the best achievable
for submodular maximization with cardinality constraints~\cite{Nemhauser1978Analysis},
it takes $O(k\cdot|U|)$ influence function evaluations for each update.
Such an inefficient update scheme makes the \emph{greedy} algorithm unable to handle a
large window size with new actions arriving at high speed.

A key challenge for efficiently supporting SIM over sliding windows
is to handle the expiry of old actions and the arrival of new actions simultaneously. 
Such a compound update pattern brings about fluctuations on users' influence
sets which potentially degrade the quality of 
previously maintained seeds. 
In the remaining of this section, 
we present a novel \emph{Influential Checkpoints} (IC) framework,
which consists of a sequence of \emph{checkpoint oracles} to efficiently handle the
expiry and the arrival of actions simultaneously. 
We first give an overview of the IC framework in Section~\ref{subsec:index}.
Then, we describe how to construct a checkpoint oracle in Section~\ref{subsec:oracle}.
Finally, we take \textsc{SieveStreaming}~\cite{Badanidiyuru2014Streaming} as an example
to illustrate the Set-Stream Mapping interface in Section~\ref{subsec:casestudy}.

\subsection{The Influential Checkpoints Framework}
\label{subsec:index}
The high level idea of the IC framework is to avoid
handling the expiry of old actions when the window slides.
Towards this goal,
the framework maintains a partial result (i.e., an influential checkpoint)
incrementally for each window slide.
When an old action expires, the outdated result is simply deleted.
In this way, the sliding window model is
transformed to a simpler \emph{append-only} model for each checkpoint,
where many existing approaches~\cite{Badanidiyuru2014Streaming,Kumar2015Fast}
can provide theoretically bounded approximate solutions.

Technically, let an influential checkpoint $\Lambda_t[i]$ $(1 \leq i\leq N)$ denote a \emph{checkpoint oracle}\footnote{ We overload the notation $\Lambda_t[i]$ to denote
the influence value returned by the oracle when it is clear in the context.} 
which provides an $\varepsilon$-approximate solution for SIM over contiguous actions
$\{W_t[i],\ldots, W_t[N]\}$.
By maintaining $N$ checkpoints (i.e., $\Lambda_t[1],\ldots,\Lambda_t[N]$),
a simple procedure to handle a window slide from $W_{t-1}$ to $W_t$
is presented in Algorithm~\ref{alg:IC}. 
Whenever a new action $a_t$ arrives, the oldest checkpoint in $W_{t-1}$ (i.e., $\Lambda_{t-1}[1]$) expires 
and a new checkpoint $\Lambda_{t}[N]$ is added to $W_t$ (Line~\ref{alg:IC-slide}).
After adding the remaining checkpoints in $W_{t-1}$ to $W_t$ (Lines~\ref{alg:IC-rename-start}-\ref{alg:IC-rename-end}), 
each checkpoint in $W_t$ processes $a_t$ as an appending action to update its partial solution
(Lines~\ref{alg:IC-update-start}-\ref{alg:IC-update-end}).
To answer the SIM query for $W_t$, we simply return the solution of $\Lambda_t[1]$.
\begin{algorithm}[ht]
    \begin{algorithmic}[1]
        \REQUIRE{IC:$\{\Lambda_{t-1}[1],\ldots,\Lambda_{t-1}[N]\}$}
        \STATE --- on receiving action $a_t$ ---
        \STATE Delete $\Lambda_{t-1}[1]$, create $\Lambda_t[N]$; \label{alg:IC-slide}
        \FORALL {$\Lambda_{t-1}[i]$} \label{alg:IC-rename-start}
            \STATE $\Lambda_t[i-1] \gets \Lambda_{t-1}[i]$;
        \ENDFOR \label{alg:IC-rename-end}
        \FORALL {$\Lambda_t[i]$} \label{alg:IC-update-start}
            \STATE $\Lambda_t[i].process(a_t)$;
        \ENDFOR \label{alg:IC-update-end}
        \STATE --- on query ---
        \RETURN the solution of $\Lambda_t[1]$;
    \end{algorithmic}
    \caption{\textsc{IC Maintenance}}
    \label{alg:IC}
\end{algorithm}

\begin{example}
    \label{exm:ic}
    Figure~\ref{fig:ic} illustrates the maintenance of checkpoints in the IC framework
    following Example~\ref{fig:example}.  
    Let $N=8$ and $k=2$. 
    The number of checkpoints always equals to the window size (i.e., $8$).
    When action $a_{10}$ arrives, $a_2$ will expire. Consequently, 
    the checkpoint $\Lambda_{10}[8]$ is created and $\Lambda_9[1]$ is deleted.
    When the SIM query is issued at time $10$,
    the result from $\Lambda_{10}[1]$ (i.e.,$\{u_2,u_3\}$) is returned.
\end{example}
It is not hard to see that once each checkpoint oracle maintains an $\varepsilon$-approximate solution 
for its \emph{append-only} action stream, IC always returns the solution with the same approximation ratio.
\subsection{Checkpoint Oracle}
\label{subsec:oracle}
The approximation ratio of IC relies on the 
checkpoint oracle providing an $\varepsilon$-approximate 
solution over an \emph{append-only} action stream. 
Although submodular maximization in an \emph{append-only} stream
has attracted many research
interests~\cite{Ausiello2012Online,Badanidiyuru2014Streaming,Saha2009Maximum,Kumar2015Fast,Yu2013SetCoverage},
they focused on a different set-stream model where elements in the stream are sets instead of actions. 
In general, an algorithm $\mathcal{A}$ on an append-only set-stream
contains two components: $f'(\cdot)$ is a monotone submodular objective function and $CX_t$
is a candidate solution containing no more than $k$ sets
from $t$ observed sets (i.e., $X_1,\ldots,X_t$).
Given a stream of sets $\{X_1,X_2,\ldots,X_m\}$,
the objective of $\mathcal{A}$ is to maximize $f'(CX_t)$ at any time $t$ ($1\leq t \leq m$).
Although this problem resembles our problem, the set-stream model cannot directly
fit in our scenario due to the following mismatch:
it strives to keep $k$ \emph{set}s from a \emph{stream of set}s
but all observed sets are immutable. 
However, SIM aims to maintain $k$ \emph{user}s from
a sequence of \emph{action}s and each arrival action may induce updates
in existing users' influence sets.

To bridge the gap between the two stream models
and leverage existing algorithms based on the set-stream model,
we propose a generic Set-Stream Mapping (SSM) interface.
The interface makes two adaptations
for a set-stream algorithm to serve as the checkpoint oracle.
First, the candidate solution $CX$ is adapted to store $k$ users.
Second, the objective function $f'$ is adapted to
the influence function $f(I_t[i] (\cdot))$,
where $I_t[i]$ denotes the influence set of user(s)
over contiguous actions $\{W_t[i],\ldots,W_t[N]\}$.
Subsequently, SSM maps an action stream to a set-stream
and feeds the set-stream to $\Lambda_t[i]$.
Whenever a new action $a_t$ arrives, the following
steps are taken for each $\Lambda_t[i]$:

\noindent(1) Identify users $u_1,u_2,\ldots,u_d$ whose $I_t[i](\cdot)$ is updated.

\noindent(2) Feed $\Lambda_t[i]$ with a stream $S'_t=\{I_t[i](u_1),\ldots,$
$I_t[i](u_d)\}$.

\noindent(3) Update the solution of $\Lambda_t[i]$ for each $I_t[i](u)\in S'_t$.

There are several choices of oracles that are developed for the set-stream model 
with differences on the solution quality, update performance, and function generality.
Typical oracles are listed in Table~\ref{tbl:oracles}.
An important conclusion is that our SSM procedure does not 
affect the quality guarantee of the mapped algorithms.
Formally:
\begin{theorem}
\label{thm:ssm}
    Let $\mathcal{A}$ be an $\varepsilon$-approximate SSO algorithm in the set-stream model
    and $\overline{\mathcal{A}}$ be the mapped algorithm of $\mathcal{A}$ using SSM.
    Then, $\overline{\mathcal{A}}$ is also
    $\varepsilon$-approximate for SIM. 
\end{theorem}
\begin{proof}
To show $\overline{\mathcal{A}}$ is also $\varepsilon$-approximate,
we consider an append-only set-stream generated by SSM over the action stream. 
At any time $t$, let $\mathsf{OPT}_t$ be the optimal influence value
of SIM wrt.~$W_t$,
and $\mathsf{OPT}^*_t$ be the optimal influence value achieved by any set of at most $k$ sets
from the mapped set-stream.
We treat all influence sets in the mapped stream as independent sets
regardless of whether they belong to the same user.
We run $\overline{\mathcal{A}}$ 
on the mapped set-stream till time $t$
and produce a result with at most $k$ sets: $CX_t=\{I_{t_a}(u_{t_a}),\ldots,I_{t_b}(u_{t_b})\}$.
Note that the influential sets in $CX_t$ may be outdated and refer to the same user.
Nevertheless, we can still use $CX_t$ to approximate $\mathsf{OPT}_t$ without affecting the approximation 
ratio. 
To obtain the seed set from $CX_t$, we select a set of distinct users $U_t$ from $CX_t$.
Since the influence function $f$ is monotone and
the up-to-date influence set of any user always grows larger in the append-only stream,
we have $f(I_t(U_t))\geq f(\cup_{X \in CX_t}X)$.
Moreover, $CX_t$ is an $\varepsilon$-approximate solution over the append-only stream,
i.e., $f(\cup_{X \in CX_t}X) \geq \varepsilon\mathsf{OPT}^*_t$. 
As the up-to-date influence sets always appear in the append-only set-stream,
we have $\mathsf{OPT}^*_t \geq \mathsf{OPT}_t$, and thus $f(I_t(U_t)) \geq \varepsilon\mathsf{OPT}_t$. 
Therefore, $U_t$ is an $\varepsilon$-approximate solution for SIM wrt.~$W_t$. 
\end{proof}

According to the \emph{SSM} steps, an action $a_t$ is mapped to at most $d$ influence sets,
where $d$ is the number of ancestors of $a_t$ in its propagation.
In practice, $d$ is usually small,
e.g., $d$ is less than $5$ on average as shown in our experiments (see Table~\ref{tbl:datasets}).
Since the number of checkpoints in the IC framework is $N$,
the total number of checkpoint evaluations is $O(dN)$.
If the update complexity of the checkpoint oracle for each set is $O(g)$,
the total time complexity of the IC framework for each action is $O(dgN)$.

In the remaining of this section,
we conduct a case study on using \textsc{SieveStreaming}~\cite{Badanidiyuru2014Streaming}
as the checkpoint oracle.
The adoption of other oracles can be similarly inferred.
\begin{table}[t]
\centering
\scriptsize
\caption{Candidate checkpoint oracles}
\label{tbl:oracles}
\begin{tabular}{|l|l|l|l|}
\hline
{\bf Oracle} & {\bf Quality} & {\bf Update} & {\bf Function}\\
\hline
\textsc{SieveStreaming}~\cite{Badanidiyuru2014Streaming} & $1/2-\beta$ & $O(\frac{\log k}{\beta})$ & General\\
\hline
\textsc{ThresholdStream}~\cite{Kumar2015Fast} & $1/2-\beta$ & $O(\frac{\log k}{\beta})$ & General\\
\hline
Blog Watch~\cite{Saha2009Maximum} & $1/4$ & $O(k)$ & Cardinality\\
\hline
M$k$C~\cite{Ausiello2012Online} & $1/4$ & $O(k\log k)$ & Cardinality\\
\hline
\end{tabular}
\end{table}

\subsection{A Case Study on SieveStreaming}
\label{subsec:casestudy}

\noindent\textbf{The SieveStreaming Algorithm}:
\textsc{SieveStreaming}~\cite{Badanidiyuru2014Streaming} works as follows:
Given a monotone submodular function $f'$
and the optimal value $\mathsf{OPT'}$ of $f'$ over the entire stream under a cardinality constraint,
\textsc{SieveStreaming} maintains a candidate solution $CX$ that includes
an incoming set $X_t$ if $CX$ has less than $k$ sets and the following holds:
\begin{equation}
\label{eq:sieve}
    f'(CX \cup \{X_t\}) - f'(CX) \geq \frac{\mathsf{OPT'}/2-f'(CX)}{k-|CX|}
\end{equation}
However, since $\mathsf{OPT'}$ is unknown in advance,
\textsc{SieveStreaming} maintains a sequence of possible values for $\mathsf{OPT'}$,
i.e., $\Omega=\{(1+\beta)^j|j \in \mathbb{Z}, m \leq (1+\beta)^j \leq 2 \cdot k \cdot m\}$
where $m=\max_X f(\{X\})$ that has been observed.
Accordingly, \textsc{SieveStreaming} keeps $|\Omega|=O(\frac{\log{k}}{\beta})$ instances to
ensure at least one of them achieves a $(1/2-\beta)$ approximation ratio
(see~\cite{Badanidiyuru2014Streaming} for more details).

\begin{figure}[t]
    \centering
    \includegraphics[width=0.475\textwidth]{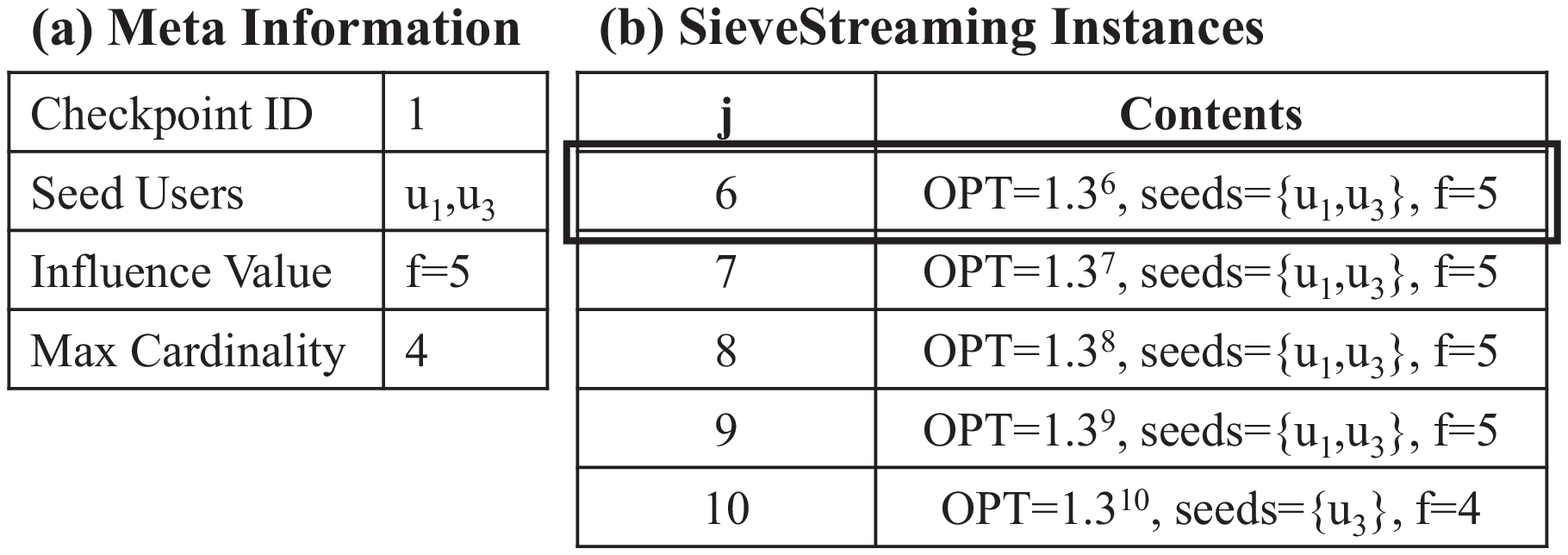}
    \caption{Contents of Checkpoint $\Lambda_8[1]$}
    \label{fig:sieve}
\end{figure}

\noindent\textbf{Set-Stream Mapping for SieveStreaming}:
Following SSM, we create $\Lambda_t[i]$ as follows:
Let $CX_i$ be the user set maintained by $\Lambda_t[i]$.
For each user $u$ with her updated influence set $I_t[i](u)$,
$\Lambda_t[i]$ selects
$u$ to $CX_i$ if $|CX_i|<k$ and:
\[
\scalebox{1.0}
{$f(I_t[i](CX_i \cup \{u\}))-f(I_t[i](CX_i))
\geq \frac{\mathsf{OPT}_{t}[i]/2-f(I_t[i](CX_i))}{k-|CX_i|}$}
\]
where $\mathsf{OPT}_{t}[i]$ is the optimal influence value achievable
on all actions from $W_t[i]$ to $W_t[N]$.
Similar to \textsc{SieveStreaming}, each $\Lambda_t[i]$ keeps a set of possible values
for $\mathsf{OPT}_{t}[i]$,
i.e., $\Omega_t[i] = \{(1+\beta)^j|j \in \mathbb{Z}, m \leq (1+\beta)^j \leq 2 \cdot k \cdot m\}$
where $m$ denotes the maximum influence value of a single influence set
over the actions $\{W_t[i],\ldots,W_t[N]\}$, i.e., $m = \max_{u\in U} f(I_t[i](u))$, 
and maintains $|\Omega_t[i]|=O(\frac{\log{k}}{\beta})$ corresponding instances. 
To answer the SIM query, we always maintain the candidate user set achieving
the largest influence value within the checkpoint.
Figure~\ref{fig:sieve} is an example for using \textsc{SieveStreaming} as the checkpoint oracle.
\begin{example}
\label{exm:sieve}
Figure~\ref{fig:sieve} illustrates the contents of $\Lambda_8[1]$ in Example~\ref{exm:ic}.
$\Lambda_8[1]$ consists of the {\it meta information} and a sequence of \textsc{SieveStreaming} instances.
In the {\it meta information}, the \emph{Checkpoint ID} indicates
the relative position of this checkpoint in the current window.
The \emph{Seed Users} and the \emph{Influence Value} are maintained
for query processing and checkpoint maintenance (as shown in Figure~\ref{fig:sieve}).
The \emph{Max Cardinality} is the maximum cardinality of a single user's influence set,
i.e., $|I_8(u_3)|=4$ for $\Lambda_8[1]$.
Suppose $\beta=0.3$, $5$ candidates with $j=6,\ldots,10$ are maintained for $\Lambda_8[1]$
($4 < 1.3^6 < \ldots < 1.3^{10} < 16$).
Each instance is maintained independently
over the mapped set-stream and the instance with the largest influence value
is used as the candidate solution (i.e., Instance with $j=6$ as highlighted in Figure~\ref{fig:sieve}).
\end{example}

Combining the results of Table~\ref{tbl:oracles} with Theorem~\ref{thm:ssm},
we can see that at least one user set maintained by $\Lambda_t[1]$
guarantees a $(1/2-\beta)$-approximate solution for SIM wrt.~$W_t$
when \textsc{SieveStreaming} is used as the checkpoint oracle.
In addition, the time complexity of IC for each update is $O(\frac{dN\log k}{\beta})$,
since the update complexity of \textsc{SieveStreaming} is $O(\frac{\log k}{\beta})$.

\section{Sparse Influential Checkpoints}
\label{sec:sic}
In the IC framework, $N$ checkpoints should be maintained to guarantee an $\varepsilon$-approximation ratio.
This implies that $O(dN)$ checkpoint oracle updates need to be performed for each arrival action.
However, real world applications often require millions of actions in one window.
Therefore, it incurs prohibitive cost to maintain all checkpoints in practice.

To reduce the number of checkpoints maintained and thus improve the update efficiency,
we design a \emph{Sparse Influential Checkpoints} (SIC)
framework to selectively maintain a subset of
checkpoints without losing too much solution accuracy as the window slides.
Specifically, the number of checkpoints maintained by SIC is logarithmic with the window size $N$
while its approximation ratio remains $\frac{\varepsilon(1-\beta)}{2}$ for any $\beta>0$
if the checkpoint oracle is $\varepsilon$-approximate.

In this section, we first present the SIC framework
and demonstrate its checkpoint maintenance in Section~\ref{subsec:sic-framework}.
In Section \ref{subsec:theoretical}, we analyze the theoretical soundness and the complexity of SIC.
Finally, we discuss how to generalize IC and SIC to handle multiple window shifts in Section~\ref{subsec:multiple}.

\subsection{The SIC Framework}
\label{subsec:sic-framework}
\begin{figure*}[ht]
    \centering
    \includegraphics[width=\textwidth]{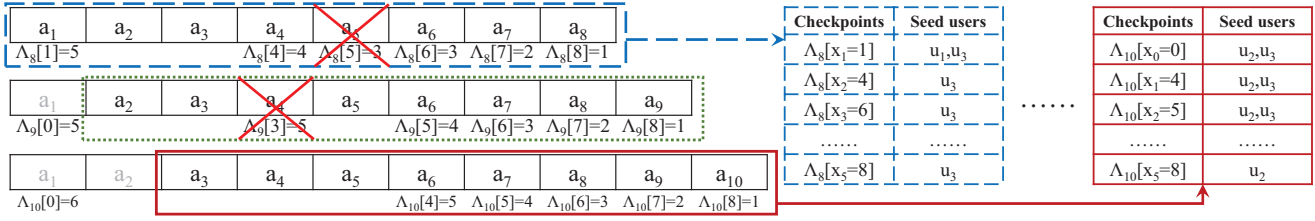}
    \caption{An example of checkpoint maintenance in SIC.}
    \label{fig:sic}
\end{figure*}
The idea of SIC is to leverage a subset of checkpoints to approximate the rest. 
On the one hand, to reduce the update cost, the number of checkpoints maintained should be as small as possible;
on the other hand, the approximation ratio should remain tight.
To achieve both goals, we propose a strategy to safely remove some checkpoints in the current window
while ensuring the remaining checkpoints are able to approximate any windows with a bounded ratio.

We consider a sequence of checkpoints
$\{\Lambda_t[x_0],\Lambda_t[x_1],$ $\ldots,$
$\Lambda_t[x_s]\}$ maintained by SIC at time $t$.
Intuitively, given any three consecutive checkpoints $\Lambda_t[x_{i-1}]$, $\Lambda_t[x_i]$, $\Lambda_t[x_{i+1}]$ kept by SIC and a parameter $\beta \in (0,1)$, 
as long as  $(1-\beta)\Lambda_t[x_{i-1}]$ is less than  $\Lambda_t[x_i]$ and $\Lambda_t[x_{i+1}]$, 
we can safely delete $\Lambda_t[x_i]$ as $\Lambda_t[x_{i+1}]$ is at least $(1-\beta)$-approximate
to $\Lambda_t[x_i]$. 
Given a checkpoint oracle with an $\varepsilon$-approximation for SIM,
it is not hard to identify that using $\Lambda_t[x_{i+1}]$ for $\mathsf{OPT}_t[x_i]$
offers an $\varepsilon(1-\beta)$ approximate solution.
Although such a maintenance strategy is simple, we need to ensure that the approximation ratio
does not degrade seriously over time,
i.e., the ratio should be at least $\frac{\varepsilon(1-\beta)}{2}$ at any time $t'>t$.
We leave this rather complex analysis to Section~\ref{subsec:theoretical}
and focus on describing the maintenance procedure in the remaining of this subsection
for ease of presentation.
\begin{algorithm}[ht]
\caption{\textsc{SIC Maintenance}}
\label{alg:SIC}
\begin{algorithmic}[1]
\REQUIRE{SIC:$\{\Lambda_{t-1}[x_0],\Lambda_{t-1}[x_1],...,\Lambda_{t-1}[x_s]\}$}
\STATE --- on receiving action $a_t$ ---
\STATE Create $\Lambda_{t}[x_{s+1}]$ where $x_{s+1}=N$;\label{alg:SIC-start}
\FORALL {$\Lambda_{t-1}[x_i]$} \label{alg:SIC-update}
    \STATE $\Lambda_t[x_i] \gets \Lambda_{t-1}[x_i]$, $x_i \gets x_i - 1$;
\ENDFOR
\FORALL {$\Lambda_t[x_i]$}
    \STATE $\Lambda_t[x_i].process(a_t)$;
\ENDFOR \label{alg:SIC-update-end}
\FORALL {$x_i$} \label{alg:SIC-delete-start}
    \STATE $\Lambda^- \gets \emptyset$;
    \FORALL {$x_j > x_i$}
        \IF{$x_{j+1}\leq x_s$ \AND $\Lambda_t[x_j] \geq (1-\beta)\Lambda_t[x_i]$
        \AND $\Lambda_t[x_{j+1}] \geq (1-\beta)\Lambda_t[x_i]$}
            \STATE $\Lambda^- \gets \Lambda^- \cup \{\Lambda_t[x_j]\}$;
        \ELSE
            \STATE \textbf{break}
        \ENDIF
    \ENDFOR
    \STATE Delete all checkpoints in $\Lambda^-$ from SIC;
    \STATE Shift the remaining checkpoints accordingly;
\ENDFOR \label{alg:SIC-delete-end}
\IF{$x_1 = 0$}\label{alg:SIC-delete-first-start}
    \STATE Delete $\Lambda_{t}[x_0]$ and shift the remaining checkpoints;
\ENDIF\label{alg:SIC-delete-first-end}
\STATE --- on query ---
\RETURN the solution of $\Lambda_t[x_1]$;
\end{algorithmic}
\end{algorithm}

Algorithm~\ref{alg:SIC} presents how to efficiently maintain the checkpoints over sliding windows in the SIC framework.
Similar to the maintenance of IC, upon receiving a new action $a_t$,
we create a new checkpoint for $a_t$ (Line~\ref{alg:SIC-start}),
add all checkpoints in $W_{t-1}$ to $W_t$,
and use $a_t$ to update all checkpoints in $W_t$ (Lines~\ref{alg:SIC-update}-\ref{alg:SIC-update-end}).
Then the efficient deletion of checkpoints are presented in Lines~\ref{alg:SIC-delete-start}-\ref{alg:SIC-delete-end}.
For each checkpoint $\Lambda_t[x_i]$,
we find the first $x_j$ ($j \geq i$) such that $\Lambda_t[x_j]\geq(1-\beta)\Lambda_t[x_i]$ and $\Lambda_t[x_{j+1}]<(1-\beta)\Lambda_t[x_i]$.
Then, all checkpoints between $x_i$ and $x_j$ are deleted and will be approximated by $\Lambda_t[x_j]$ 
in the subsequent window slides.
Finally, if the second checkpoint (i.e., $\Lambda_{t}[x_1]$) has expired,
the earliest checkpoint (i.e., $\Lambda_{t}[x_0]$) will be deleted
(Lines~\ref{alg:SIC-delete-first-start}-\ref{alg:SIC-delete-first-end}).
It is notable that an additional checkpoint ($\Lambda_t[x_0]$) is stored in SIC to keep track of
the solution over a window with size larger than $N$.
Since $\Lambda_t[x_0]$ approximates the upper bound of the optimal solution for the current window 
and Algorithm~\ref{alg:SIC} always maintains a bounded ratio between two neighboring checkpoints,
a bounded approximation ratio is guaranteed by using $\Lambda_t[x_1]$ as the solution for the current window.
Figure~\ref{fig:sic} and Example~\ref{exm:sic} provide the running example of the SIC maintenance.
\begin{example}
\label{exm:sic}
    Figure~\ref{fig:sic} illustrates the maintenance of checkpoints in the SIC framework following
    Example~\ref{exm:def}. 
    Let $N=8$, $k=2$ and $\beta=0.3$. There are initially $6$ checkpoints in SIC at time $8$. 
    According to Algorithm~\ref{alg:SIC},
    $\Lambda_8[5]$ is deleted from SIC since $\Lambda_8[6] = 3 > (1-0.3)\times 3 = (1-\beta)\Lambda_8[4]$. 
    At time $8$, $\Lambda_8[1]$ will be used to answer the SIM query.
    As the window shifts at time $9$ with the arrival of $a_9$,
    $a_1$ and $\Lambda_8[1]$ (which later becomes $\Lambda_9[0]$) expire.
    But $\Lambda_9[0]$ is stored because $\Lambda_9[3]$ has not expired yet.
    Then all checkpoints will be updated according to $a_9$.
    After the update procedure, we find $\Lambda_9[3]$ can be deleted since
    $\Lambda_9[5] > (1-\beta)\Lambda_9[0]$.
    Finally, all checkpoints are updated according to $a_{10}$ at time $10$
    and no checkpoints are deleted.
    $\Lambda_{10}[4]$ will be used to answer the SIM query at time $10$.
\end{example}

In the following, we will demonstrate the theoretical soundness of SIC and also analyze the complexity of SIC.

\subsection{Theoretical Analysis}
\label{subsec:theoretical}
To establish our theoretical claims for SIC, we first analyze the property 
of the optimal checkpoint oracle which always returns the optimal solution for SIM
over an append-only action stream.
There are two important properties of the optimal checkpoint oracle. 

\begin{definition}[Monotonicity \& Subadditivity]\ \\
    Let $t_a\leq t_b$ be two timestamps and $W^{t_a}_{t_b}$ 
    represents a window containing a set of contiguous actions: $a_{t_a},...,a_{t_b}$
    with the corresponding checkpoint denoted as $\Lambda^{t_a}_{t_b}$. 
    Given any $t_1,t_2,t_3$ s.t. $t_1\leq t_2\leq t_3$,
    the checkpoint is monotone if $\Lambda^{t_1}_{t_3} \geq \Lambda^{t_1}_{t_2}$.
    Moreover, the checkpoint is subadditive if
    $\Lambda^{t_1}_{t_3} \leq \Lambda^{t_1}_{t_2} + \Lambda^{t_2}_{t_3}$. 
\end{definition}

\begin{lemma}\label{lm:optimal}
    Let $t_a\leq t_b$ be two timestamps and $\mathsf{OPT}^{t_a}_{t_b}$ denote the optimal oracle
    (as well as the optimal value) for $W^{t_a}_{t_b}$.
    The optimal oracle is both monotone and subadditive. 
\end{lemma}
\begin{proof}
    Let $I^{t_a}_{t_b}(S)$ be the influence set of $S$
    and $S^{opt}_{t_a,t_b}$ be the optimal solution of SIM
    for contiguous actions $\{W_{t_a},$ $\ldots,$ $W_{t_b}\}$.
    Because $S^{opt}_{t_1,t_2}$ must be a candidate solution for $W^{t_1}_{t_3}$,
    it is obvious to see $\mathsf{OPT}^{t_1}_{t_3} \geq \mathsf{OPT}^{t_1}_{t_2}$.
    Then, we prove $\mathsf{OPT}^{t_1}_{t_3} \leq \mathsf{OPT}^{t_1}_{t_2} + \mathsf{OPT}^{t_2}_{t_3}$.
    For any $S \subseteq U$, we have $I^{t_1}_{t_3}(S) = I^{t_1}_{t_2}(S) \cup I^{t_2}_{t_3}(S)$.
    Then, we have:
    \begin{align*}
        \mathsf{OPT}^{t_1}_{t_3} = f(S^{opt}_{t_1,t_3})
        & \leq f(I^{t_1}_{t_2}(S^{opt}_{t_1,t_3})) + f(I^{t_2}_{t_3}(S^{opt}_{t_1,t_3})) \\
        & \leq \mathsf{OPT}^{t_1}_{t_2} + \mathsf{OPT}^{t_2}_{t_3}
    \end{align*}
    where the first inequality holds for the property of influence sets,
    and the second inequality is satisfied because $S^{opt}_{t_1,t_3}$
    is also a candidate solution for both~$W^{t_1}_{t_2}$ and~$W^{t_2}_{t_3}$.
\end{proof}

We note that although the optimal checkpoint oracle is both monotone and subadditive,
it is intractable unless $P=NP$.
In practice, we utilize the approximate checkpoint oracles as listed in Table~\ref{tbl:oracles}.
The approximate oracles are monotone. 
This is essential due to their greedy nature: updating the maintained result only
when this update increases the function value.
Given the monotonicity of the approximate checkpoint oracles, 
the monotonicity and subadditivity of the optimal oracle,
we are ready to prove that the checkpoint maintenance strategy used in SIC is theoretically bounded.
\begin{lemma}
\label{lm:smoothness}
    Given any $t_1,t_2,t_3,t_4$ s.t. $t_1\leq t_2\leq t_3\leq t_4$, 
    $\forall \beta \in (0,1)$, if $(1-\beta)\Lambda^{t_1}_{t_3} \leq \Lambda^{t_2}_{t_3}$,
    then $\frac{\varepsilon(1-\beta)}{2}\mathsf{OPT}^{t_1}_{t_4} \leq \Lambda^{t_2}_{t_4}$.
\end{lemma}
\begin{proof}
    The following inequalities hold:
    \begin{align*}
    \Lambda^{t_2}_{t_4} & \geq \frac{1}{2}(\Lambda^{t_2}_{t_3} +\Lambda^{t_2}_{t_4})
    \geq \frac{1}{2}((1-\beta)\Lambda^{t_1}_{t_3}+\Lambda^{t_2}_{t_4}) \\
    & \geq \frac{1-\beta}{2}(\Lambda^{t_1}_{t_3} + \Lambda^{t_2}_{t_4})
    \geq \frac{\varepsilon(1-\beta)}{2}(\mathsf{OPT}^{t_1}_{t_3}+\mathsf{OPT}^{t_2}_{t_4}) \\
    & \geq \frac{\varepsilon(1-\beta)}{2}\mathsf{OPT}^{t_1}_{t_4}
    \end{align*}
    where the first inequality holds from the monotonicity of the approximate checkpoint oracles;
    the second inequality is due to the condition that $\Lambda_{t_3}^{t_2} \geq (1-\beta)\Lambda^{t_1}_{t_3}$;
    the third inequality is obvious since $\beta\in(0,1)$;
    the fourth inequality holds because of the approximation ratio of checkpoint oracles and 
    the final inequality holds as the optimal checkpoint oracle is both monotone and subadditive.
\end{proof}

According to Lemma~\ref{lm:smoothness},
if $(1-\beta)\Lambda^{t_1}_{t_3} \leq \Lambda^{t_2}_{t_3}$,
using the checkpoint oracle started at $t_2$ to approximate any checkpoints between $t_1$ and $t_2$
always achieves an $\frac{\varepsilon(1-\beta)}{2}$ approximation
for any number of appending actions.
Next, we present Lemma~\ref{lm:SIC-property} to demonstrate the property of the checkpoints maintained by Algorithm~\ref{alg:SIC}.

\begin{lemma}
\label{lm:SIC-property}
    The SIC on window $W_t$ contains
    $s$ checkpoints $\Lambda_t[x_0],\Lambda_t[x_1],\ldots,\Lambda_t[x_s]$ $(x_0<x_1<\ldots<x_s)$
    maintained by Algorithm~\ref{alg:SIC}.
    Given a constant $\beta \in (0,1)$,
    any neighboring checkpoints $\Lambda_t[x_i]$, $\Lambda_t[x_{i+1}]$ and $\Lambda_t[x_{i+2}]$
    satisfy one of the following conditions:
    \begin{enumerate}[noitemsep,leftmargin=*]
        \item{if $\Lambda_t[x_{i+1}] \geq (1-\beta)\Lambda_t[x_i]$, then $\Lambda_t[x_{i+2}]<(1-\beta)\Lambda_t[x_i]$.}
        \item{if $x_{i+1} \neq x_i + 1 \wedge \Lambda_t[x_{i+1}]<(1-\beta)\Lambda_t[x_i]$,
        then $\frac{\varepsilon(1-\beta)}{2}\\\mathsf{OPT}_t[x_i] \leq \Lambda_t[x_{i+1}]$.}
        \item{$x_{i+1} = x_i + 1 \wedge \Lambda_t[x_{i+1}] < (1-\beta)\Lambda_t[x_i]$.}
    \end{enumerate}
\end{lemma}

\begin{proof}
    We prove the lemma by induction. 
    As the base case, initially there are only 2 actions in the window and either condition 1 or 3 holds.
    
    Next, assume Lemma~\ref{lm:SIC-property} holds at time $t$ and 
    we show that it still holds after the update procedure in Algorithm~\ref{alg:SIC} at time $t+1$. 
    Let $\Lambda_{t}[x_i]$ be a checkpoint instantiated before $t+1$ and is not deleted during the update procedure at $t+1$,
    then $\Lambda_t[x_{i+1}]$ is the subsequent checkpoint of $\Lambda_{t+1}[x_i]$ at time $t$,
    Next, we discuss all possible cases when performing the update procedure of Algorithm~\ref{alg:SIC} at time $t+1$:
    
    \noindent\textbf{Case 1}: $x_{i+1} \neq x_i + 1$ and $\Lambda_{t}[x_{i+1}]$ is deleted at $t+1$.
    In this case, we have $\Lambda_{t+1}[x_{i+1}] \geq (1-\beta)\Lambda_{t+1}[x_i]$
    and $\Lambda_{t+1}[x_{i+2}]<(1-\beta)\Lambda_{t+1}[x_i]$
    according to Lines~\ref{alg:SIC-delete-start}-\ref{alg:SIC-delete-end} of Algorithm~\ref{alg:SIC}.
    In this case, condition 1 holds at $t+1$.
    
    \noindent\textbf{Case 2}: $x_{i+1} \neq x_i + 1$ and $\Lambda_{t+1}[x_{i+1}]$ is not deleted at $t+1$. 
    In this case, $\Lambda_{t+1}[x_{i+1}]$ must become the subsequent checkpoint of $\Lambda_{t+1}[x_i]$ at some time $t'\leq t$. 
    Then, at $t'$, we have $\Lambda_{t'}[x_{i+1}] \geq (1-\beta)\Lambda_{t'}[x_i]$. 
    According to Lemma~\ref{lm:smoothness},
    $\Lambda_{t+1}[x_{i+1}]\geq\frac{\varepsilon(1-\beta)}{2}\mathsf{OPT}_{t+1}[x_i]$ holds. 
    Because $\Lambda_{t+1}[x_{i+1}]$ is not deleted at $t+1$,
    we have either condition 1 (when $\Lambda_{t+1}[x_{i+1}] \geq (1-\beta)\Lambda_{t+1}[x_i]$) 
    or condition 2 holds (when $\Lambda_{t+1}[x_{i+1}] < (1-\beta)\Lambda_{t+1}[x_i]$) at $t+1$. 
    
    \noindent\textbf{Case 3}: $x_{i+1}=x_i + 1$. If $\Lambda_{t+1}[x_{i+1}] \geq (1-\beta)\Lambda_{t+1}[x_i]$, 
    then condition 1 holds since $\Lambda_{t+1}[x_{i+1}]$ is not deleted at $t+1$; otherwise, condition 3 holds.
    
    Therefore, at least one condition in Lemma~\ref{lm:SIC-property} holds in all possible cases at $t+1$
    and we conclude the proof. 
\end{proof}

Leveraging Lemma~\ref{lm:SIC-property}, we are able to analyze SIC theoretically.
First, from conditions 1 and 2, we can infer that if there are checkpoints deleted between $x_i$ and $x_{i+1}$, 
the ratios between $\Lambda_t[x_{i+1}]$ and the optimal solution of deleted checkpoints
are guaranteed to be at least $\frac{\varepsilon(1-\beta)}{2}$.
Next, by collectively examining conditions 1--3,
we can see that there is at least one checkpoint in $\Lambda_t[x_{i+1}]$ and $\Lambda_t[x_{i+2}]$
returning an influence value of smaller than $(1-\beta)\Lambda_t[x_i]$,
and thus the number of checkpoints maintained is $O(\frac{\log N}{\beta})$.
Based on these intuitions, we then formally state the approximation guarantee and the complexity of SIC
in Theorems~\ref{thm:ratio}--\ref{thm:complexity}:

\begin{theorem}
\label{thm:ratio}
    SIC maintains a $\frac{\varepsilon(1-\beta)}{2}$-approximate solution
    for SIM in $\Lambda_t[x_1]$ when a $\varepsilon$-approximate checkpoint oracle is used.
\end{theorem}
\begin{proof}
    We use $\mathsf{OPT}_t$ to denote the optimal solution of SIM w.r.t. $W_t$
    and we prove that $\frac{\varepsilon(1-\beta)}{2}$ is a lower bound for
    the ratio between $\Lambda_t[x_1]$ and $\mathsf{OPT}_t$. 
    
    Let $\Lambda_t[x_0]$ be the expired checkpoint just before $\Lambda_t[x_1]$.
    Since $\Lambda_t[x_0]$ and $\Lambda_t[x_1]$ are neighboring checkpoints in SIC, 
    one of the conditions in Lemma~\ref{lm:SIC-property} holds at time $t$.
    
    If condition 3 in Lemma~\ref{lm:SIC-property} holds, we have $\mathsf{OPT}_t \leq \varepsilon \Lambda_t[x_1]$
    since $\Lambda_t[x_1]$ directly maintains an approximate solution on $W_t$.
    Otherwise, we have:
    $\mathsf{OPT}_t \leq \mathsf{OPT}_t[x_0] \leq \frac{2}{\varepsilon (1-\beta)}\Lambda_t[x_1]$
    since $\Lambda_t[x_0]$ has expired.
    Thus, SIC maintains an at least $\frac{\varepsilon(1-\beta)}{2}$-approximate solution in $\Lambda_t[x_1]$.
\end{proof}
\begin{theorem}
    \label{thm:sieve-ratio}
    SIC obtains a $(1/4-\beta)$-approximate solution
    for SIM when \textsc{SieveStreaming} is used as the checkpoint oracle. 
\end{theorem}
\begin{proof}
    Since the \textsc{SieveStreaming} algorithm guarantees a $(1/2-\beta)$ approximation ratio to the optimal solution, 
    SIC with \textsc{SieveStreaming} as the checkpoint oracle
    preserves a $\frac{1}{2}(\frac{1}{2}-\beta)(1-\beta)$ approximation guarantee
    according to Theorem~\ref{thm:ratio}.
    As $\frac{1}{2}(\frac{1}{2}-\beta)(1-\beta)=\frac{1}{4}-\frac{3}{4}\beta+\beta^2>\frac{1}{4}-\beta$, 
    we get at least a $(1/4-\beta)$-approximate solution for SIM.
\end{proof}
\begin{theorem}
\label{thm:complexity}
    The number of checkpoints maintained by SIC wrt.~a sliding window of size $N$ is $O(\frac{\log N}{\beta})$. 
\end{theorem}
\begin{proof}
    Lemma~\ref{lm:SIC-property} guarantees either $\Lambda_t[x_{i+1}]$ or $\Lambda_t[x_{i+2}]$ is less than $(1-\beta)\Lambda_t[x_i]$. 
    Since we have $\Lambda_t[x_1]/\Lambda_t[N]$ is bounded by $O(N)$,
    the number of checkpoints is at most $\frac{2\cdot\log{N}}{\log (1-\beta)^{-1}}$ for $\beta \in (0,1)$. 
    Therefore, the number of checkpoints maintained by SIC is $O(\frac{\log N}{\beta})$.
\end{proof}

As the time complexity for a checkpoint to update each action is
$O(dg)$ if each checkpoint takes $O(g)$ to evaluate one influence set
and the number of checkpoints maintained by SIC is $O(\frac{\log N}{\beta})$,
the time complexity of SIC to update each action
is $O(\frac{dg\log N}{\beta})$.
When \textsc{SieveStreaming} is used as the checkpoint oracle, we have $g=O(\frac{\log k}{\beta})$
and thus the time complexity of SIC for each update
is $O(\frac{d\log N\log k}{\beta^2})$.

\subsection{Handling Multiple Window Shifts}
\label{subsec:multiple}
Although we have discussed how to handle SIM queries 
for windows which shift for one action at a time,
many applications do not require to retrieve the result at such an intense rate.
Hereby, we discuss how to handle multiple window shifts,
i.e., each window shift receives $L$ new actions
while the earliest $L$ actions become expired at the same time. 

To handle multiple window shifts for IC,
we create only one new checkpoint and delete the earliest checkpoint
when the window slides from time $t$ to $t+L$.
Subsequently, all actions from $a_{t+1}$ to $a_{t+L}$ are collected to update all checkpoints in the window.
Thus, the number of checkpoints created for multiple window
shifts will be $\left\lceil\frac{N}{L}\right\rceil$.
On top of the IC maintenance strategy, we still use the same SIC algorithm over the checkpoints created 
by IC to support multiple window shifts. 

Lastly, the aforementioned maintenance strategies still preserve the theoretical results
as there is no fundamental differences between handling single window shift and multiple window shifts using 
our proposed frameworks.

\section{Experimental Results}
\label{sec:experiments}

In this section, we evaluate the efficiency and effectiveness of our proposed frameworks
on several real-world and synthetic datasets.
First, we compare IC and SIC for influence values and processing efficiency with varying $\beta$.
Then, we compare the solution qualities and throughputs of all approaches
with different seed set size.
Finally, we evaluate the scalability of all compared approaches.

\subsection{Experimental Setup}
\label{subsec:setup}
\noindent\textbf{Datasets}:
We collect two real-world datasets and synthesize two datasets for extensive studies.
\begin{itemize}[nosep,leftmargin=*]
    \item{
        \textbf{Reddit}:
        Reddit is an online forum where user actions include \emph{post} and \emph{comment}.
        We collect all Reddit \emph{comment} actions in May 2015 from
        \emph{kaggle}\footnote{\url{https://www.kaggle.com/reddit/reddit-comments-may-2015}}
        and query the Reddit API for the \emph{post} actions in the same period.
        The dataset contains $48,104,875$ actions from $2,628,904$ users.
    }
    \item{
        \textbf{Twitter}:
        Twitter is an online social network where actions
        include \emph{tweet}, \emph{retweet}, \emph{quote} and \emph{reply}.
        We crawl these actions for one week via Twitter stream API\footnote{\url{https://dev.twitter.com/streaming/overview}}
        on trending topics such as US presidential election, 2016 NBA finals and Euro 2016.
        The dataset contains $9,724,908$ actions from $2,881,154$ users.
    }
    \item{
        \textbf{Synthetic Datasets}:
        We synthesize two action streams with different response patterns
        to test the robustness of the proposed solutions.
        There are two types of actions in concern: \emph{post} and \emph{follow}.
        We use the R-MAT model~\cite{Chakrabarti2004RMAT} to synthesize 5 different power law graphs
        with the number of users ranging from $1$-$5$ million ($2$ million by default).
        For each synthetic graph, we generate $10$ million actions
        by randomly selecting a user to perform either a \emph{post} or a \emph{follow} action.
        If an action $a_t$ is \emph{follow},
        it will respond to a previous action $a_{t'}$ with a response distance $\Delta=t-t'$.
        To demonstrate different response patterns, two datasets are generated based on the distances
        conforming to exponential distributions with different parameters:
        (1)\textbf{SYN-O}: $\Delta\sim\mathtt{exp}(\lambda=2.0\times 10^{-6})$,
        which indicates ``old posts get more followers'';
        (2)\textbf{SYN-N}: $\Delta\sim\mathtt{exp}(\lambda=2.0\times 10^{-4})$,
        which represents ``recent posts get more followers''.
    }
\end{itemize}
The statistics of these datasets are summarized in Table~\ref{tbl:datasets}.
\begin{table}[t]
    \setlength\tabcolsep{2pt}
    \small
    \centering
    \caption{Statistics on datasets}
    \label{tbl:datasets}
    \begin{tabular}{|c|c|c|c|c|} \hline 
        {\bf Dataset} & {\bf Users} & {\bf Actions} & {\bf Resp. dist.} & {\bf Avg. depth} \\ \hline
        Reddit & 2,628,904 & 48,104,875 & 404714.9 & 4.58 \\ \hline
        Twitter & 2,881,154 & 9,724,908 & 294609.4 & 1.87 \\ \hline
        SYN-O & 1M--5M & 10,000,000 & 500,000 & 2.5 \\ \hline
        SYN-N & 1M--5M & 10,000,000 & 5,000 & 2.59 \\ \hline
    \end{tabular}
\end{table}

\begin{figure*}[htpb]
    \captionsetup[subfloat]{farskip=1pt,captionskip=1pt}
    \centering
    \subfloat[Reddit]{%
        \includegraphics[width=0.235\textwidth]{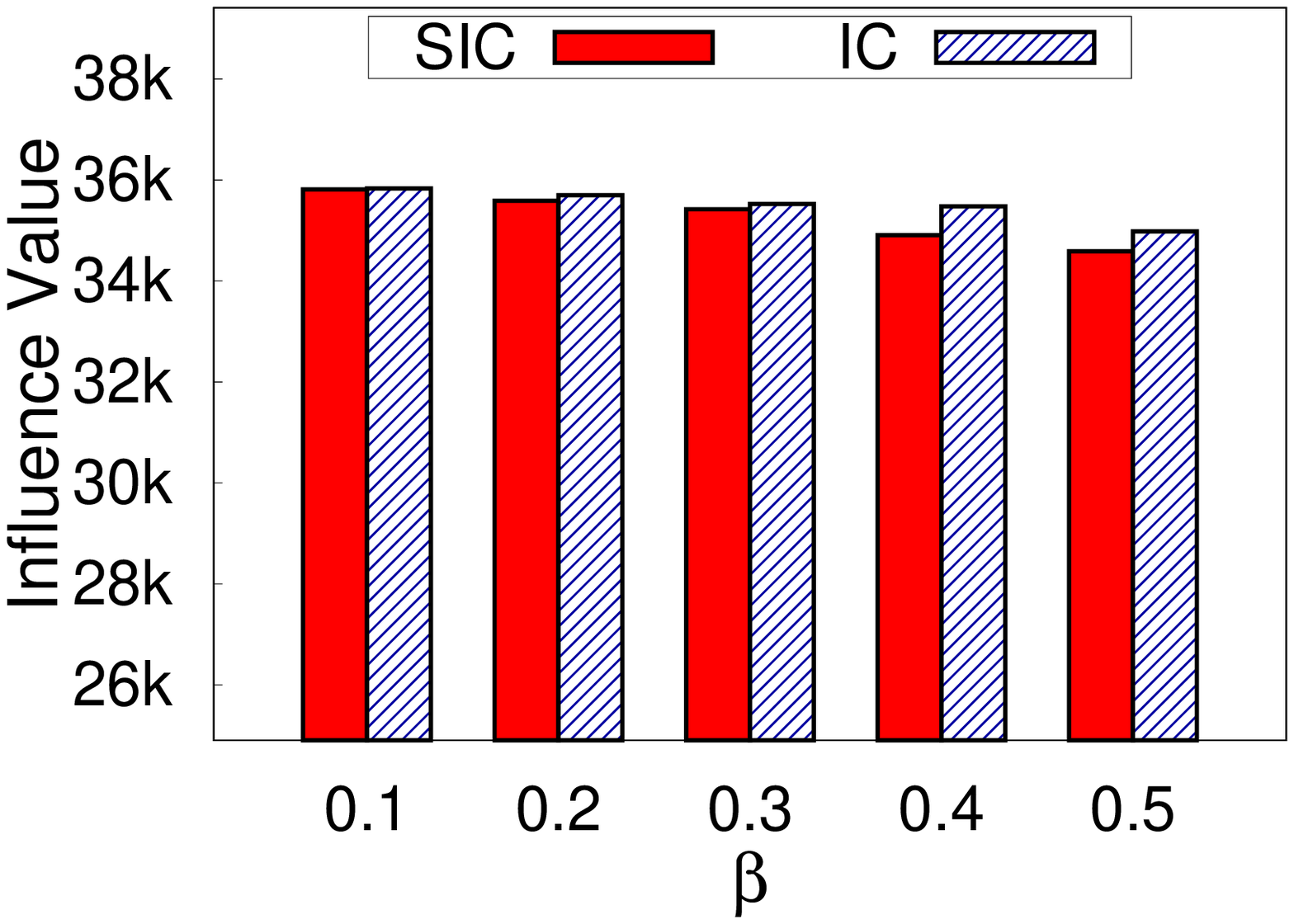}
        \label{fig:reddit-e-quality}
    }
    \subfloat[Twitter]{%
        \includegraphics[width=0.235\textwidth]{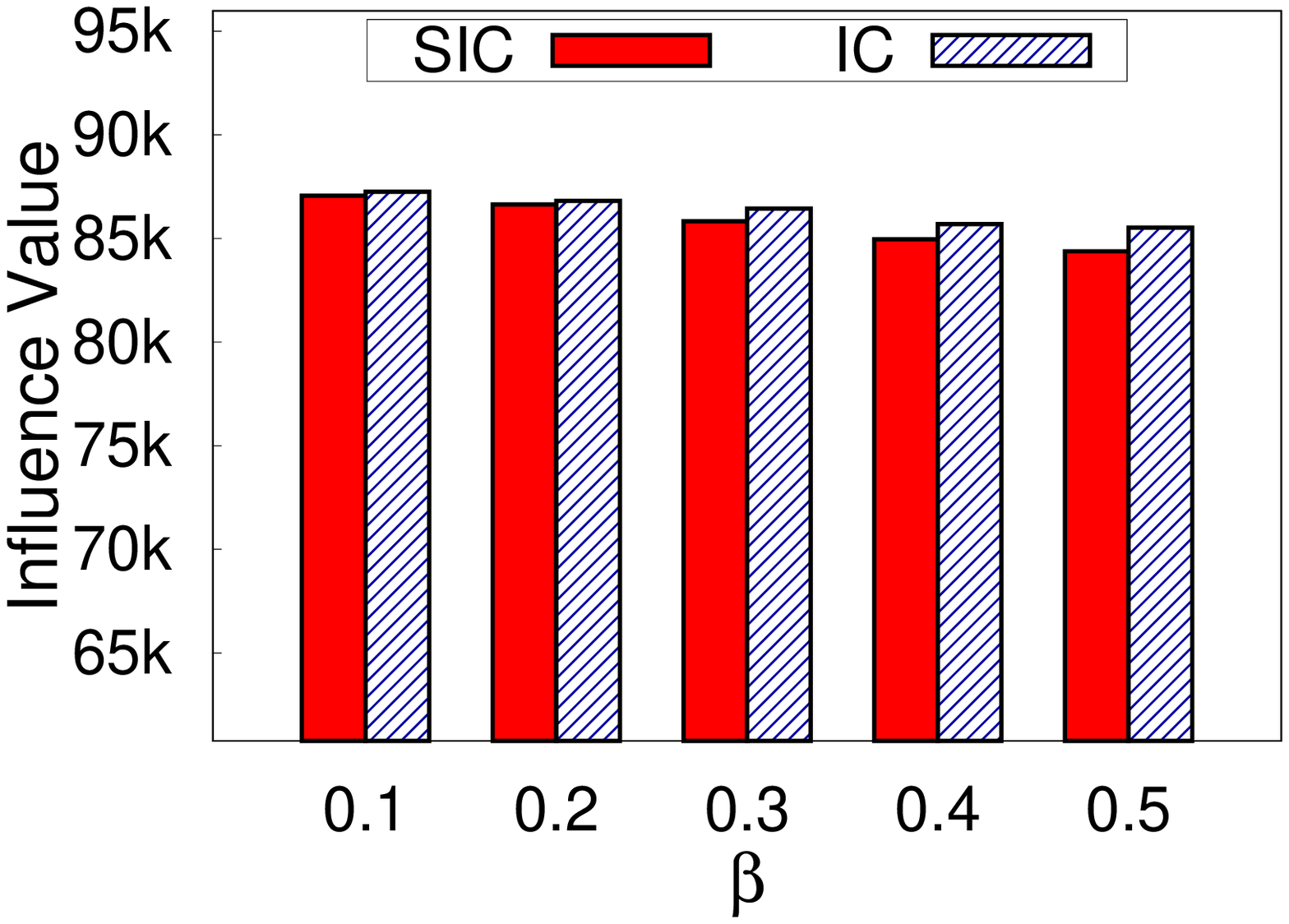}
        \label{fig:twitter-e-quality}
    }
    \subfloat[SYN-O]{%
        \includegraphics[width=0.235\textwidth]{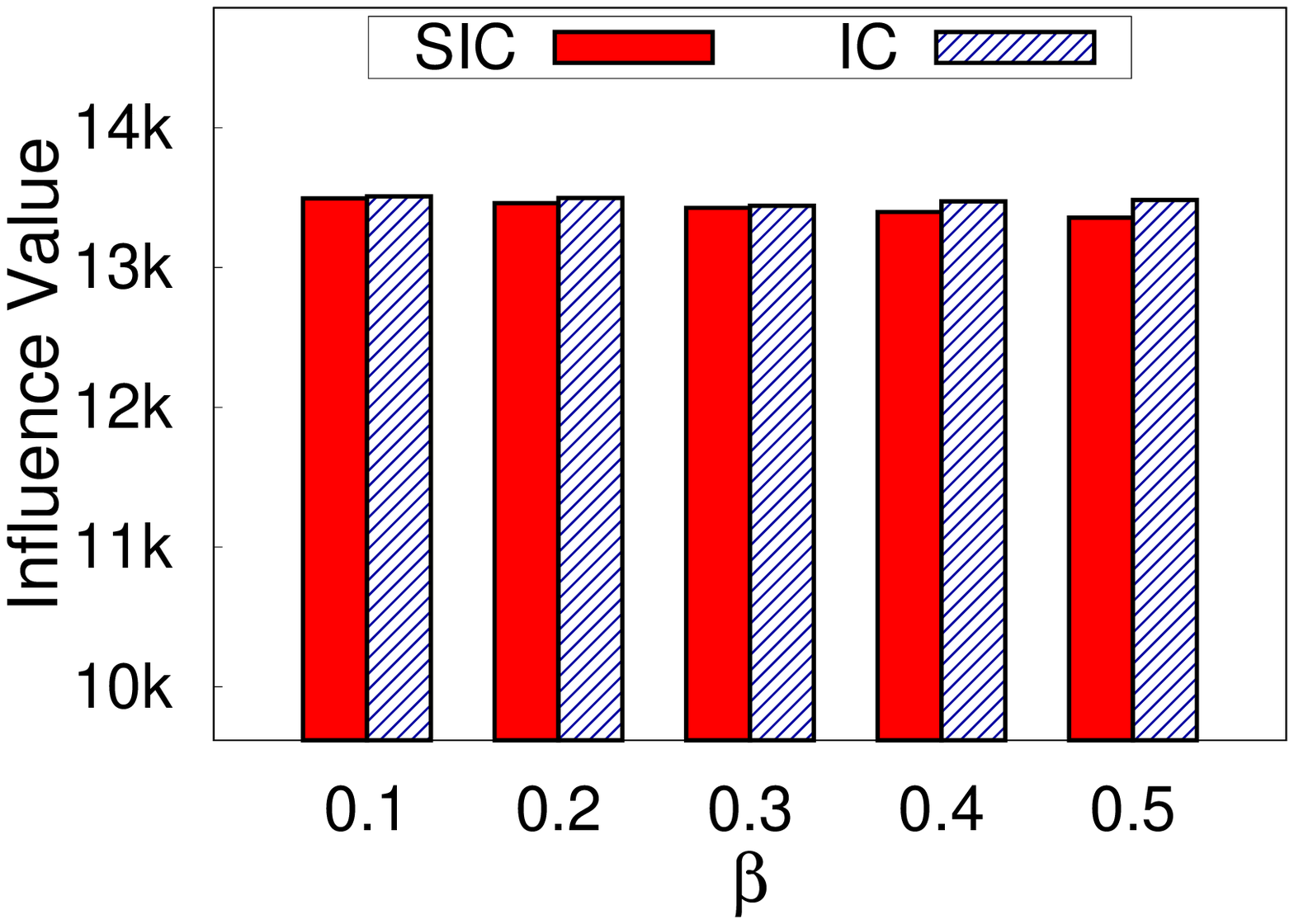}
        \label{fig:syn-o-e-quality}
    }
    \subfloat[SYN-N]{%
        \includegraphics[width=0.235\textwidth]{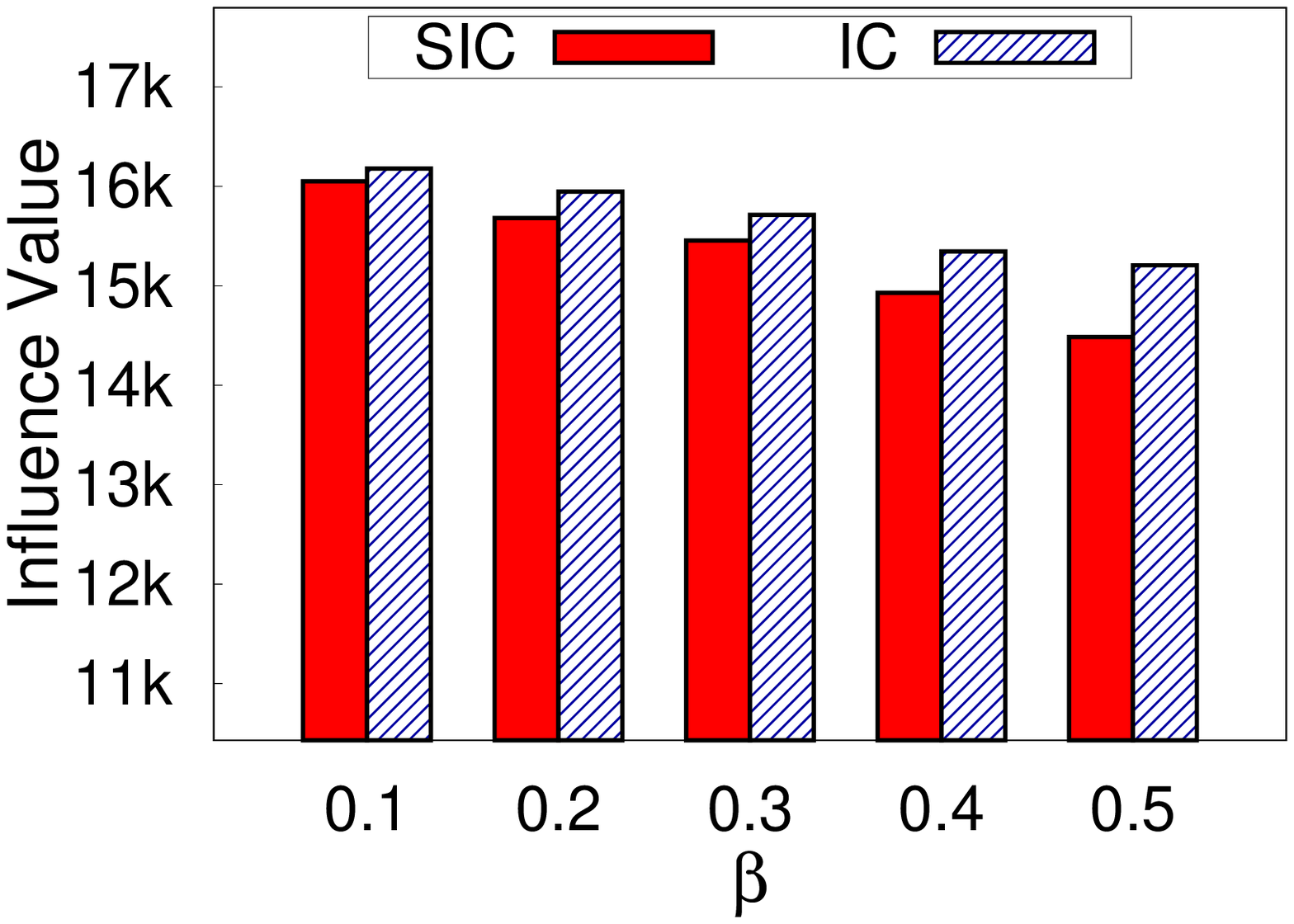}
        \label{fig:syn-n-e-quality}
    }
    \caption{Influence Values of IC and SIC with varying $\beta$.}
    \label{fig:qualitiesBeta}
\end{figure*}
\begin{figure*}[htpb]
    \captionsetup[subfloat]{farskip=1pt,captionskip=1pt}
    \centering
    \subfloat[Reddit]{%
        \includegraphics[width=0.235\textwidth]{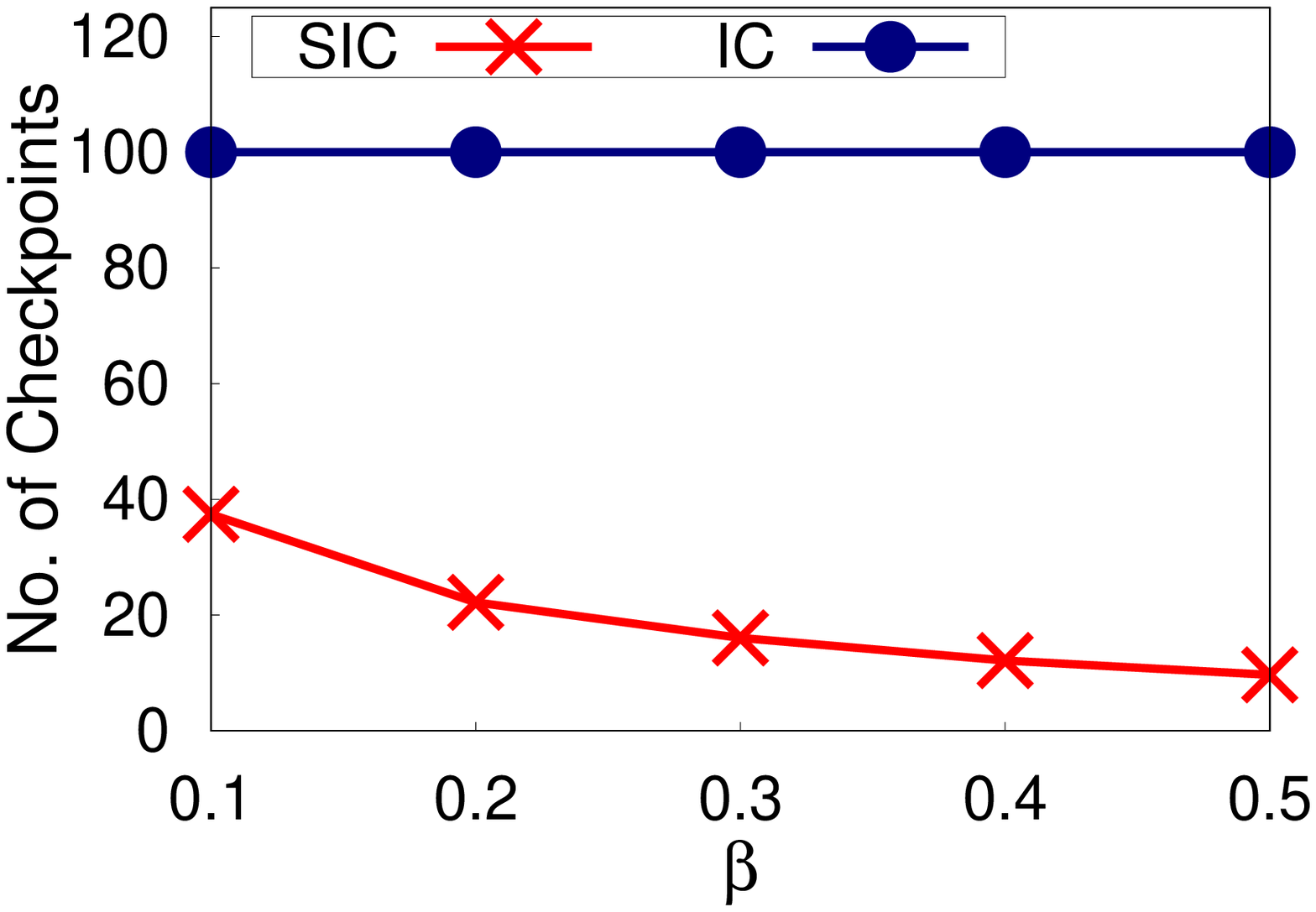}
        \label{fig:reddit-e-checkpoints}
    }
    \subfloat[Twitter]{%
        \includegraphics[width=0.235\textwidth]{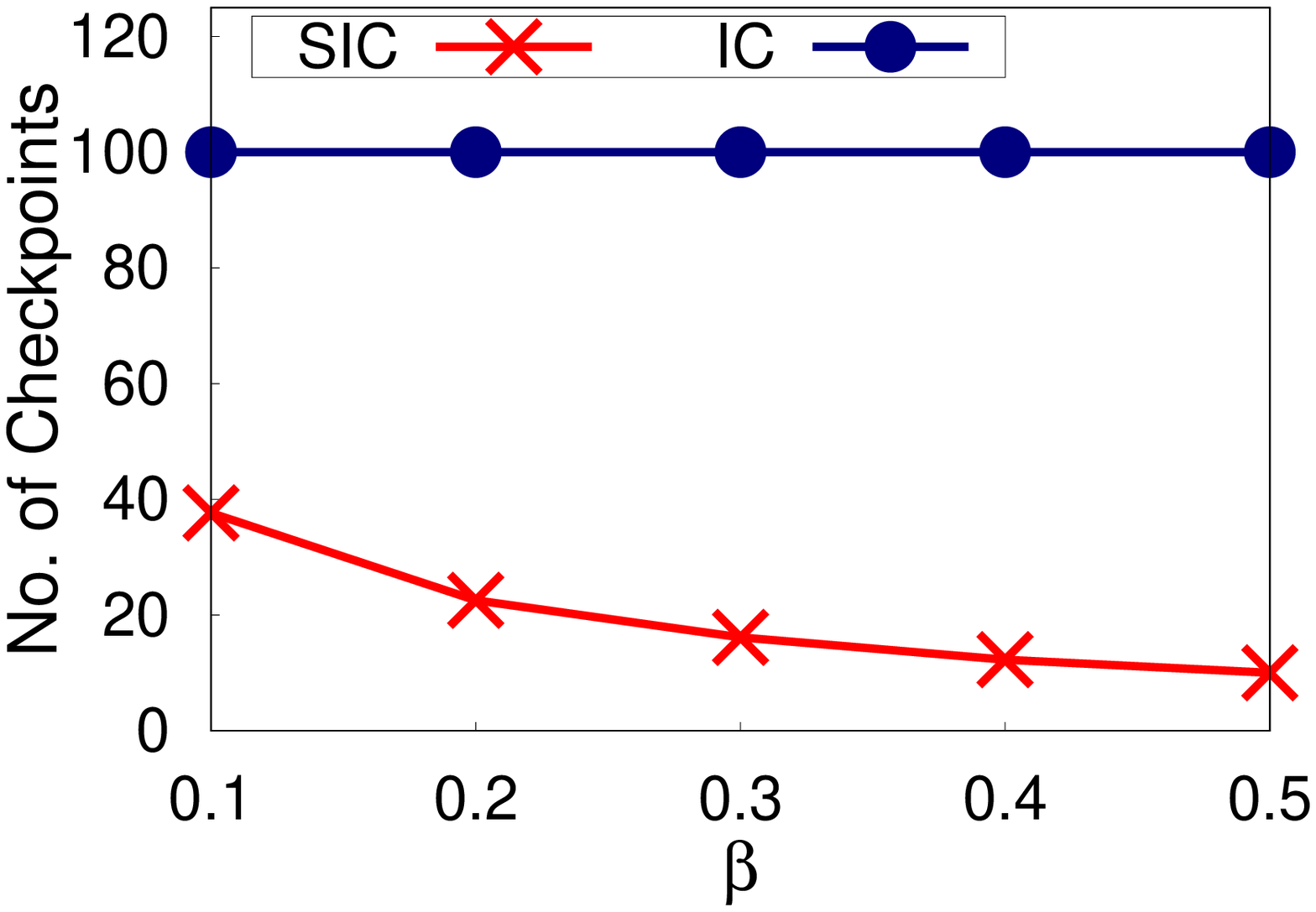}
        \label{fig:twitter-e-checkpoints}
    }
    \subfloat[SYN-O]{%
        \includegraphics[width=0.235\textwidth]{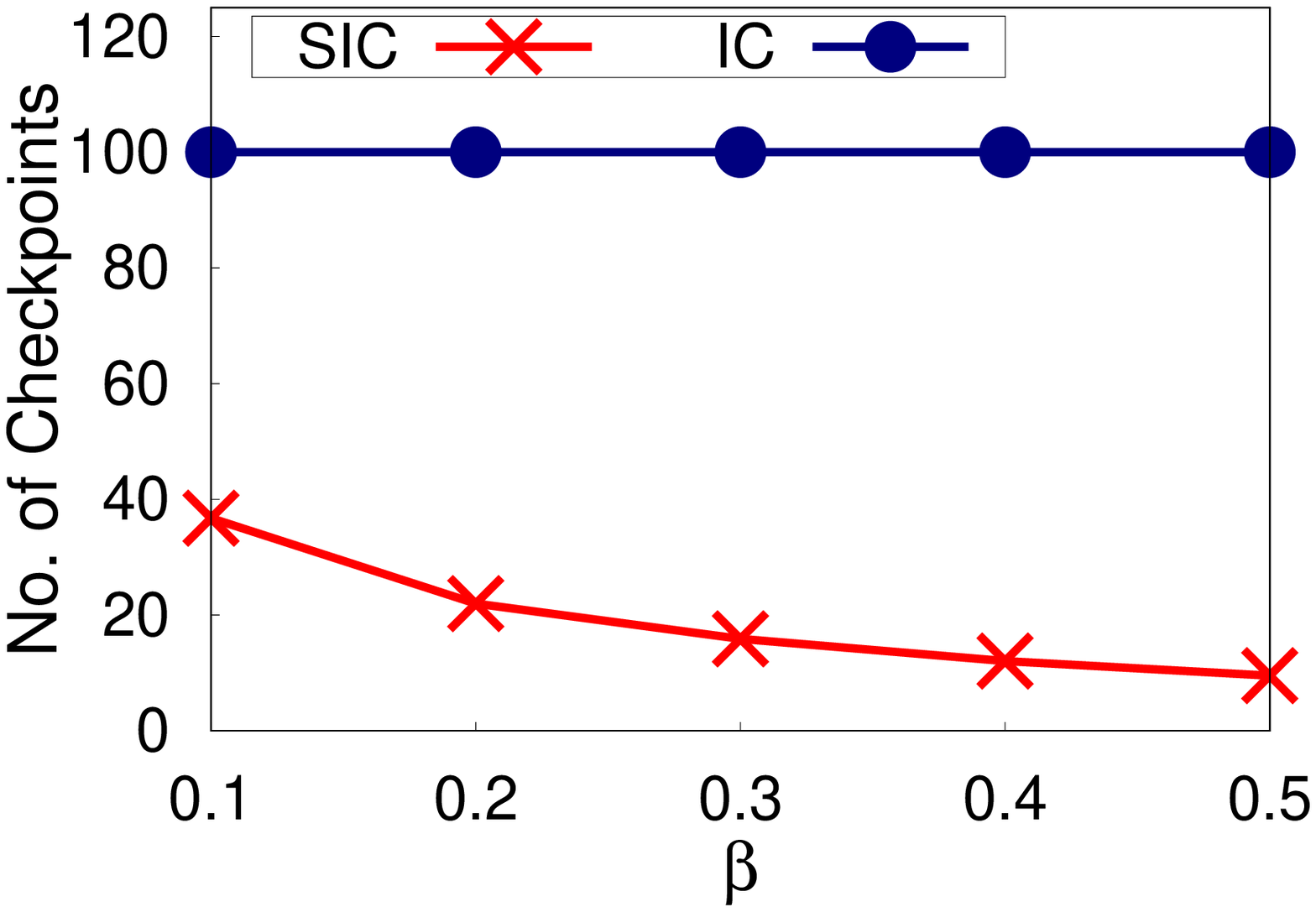}
        \label{fig:syn-o-e-checkpoints}
    }
    \subfloat[SYN-N]{%
        \includegraphics[width=0.235\textwidth]{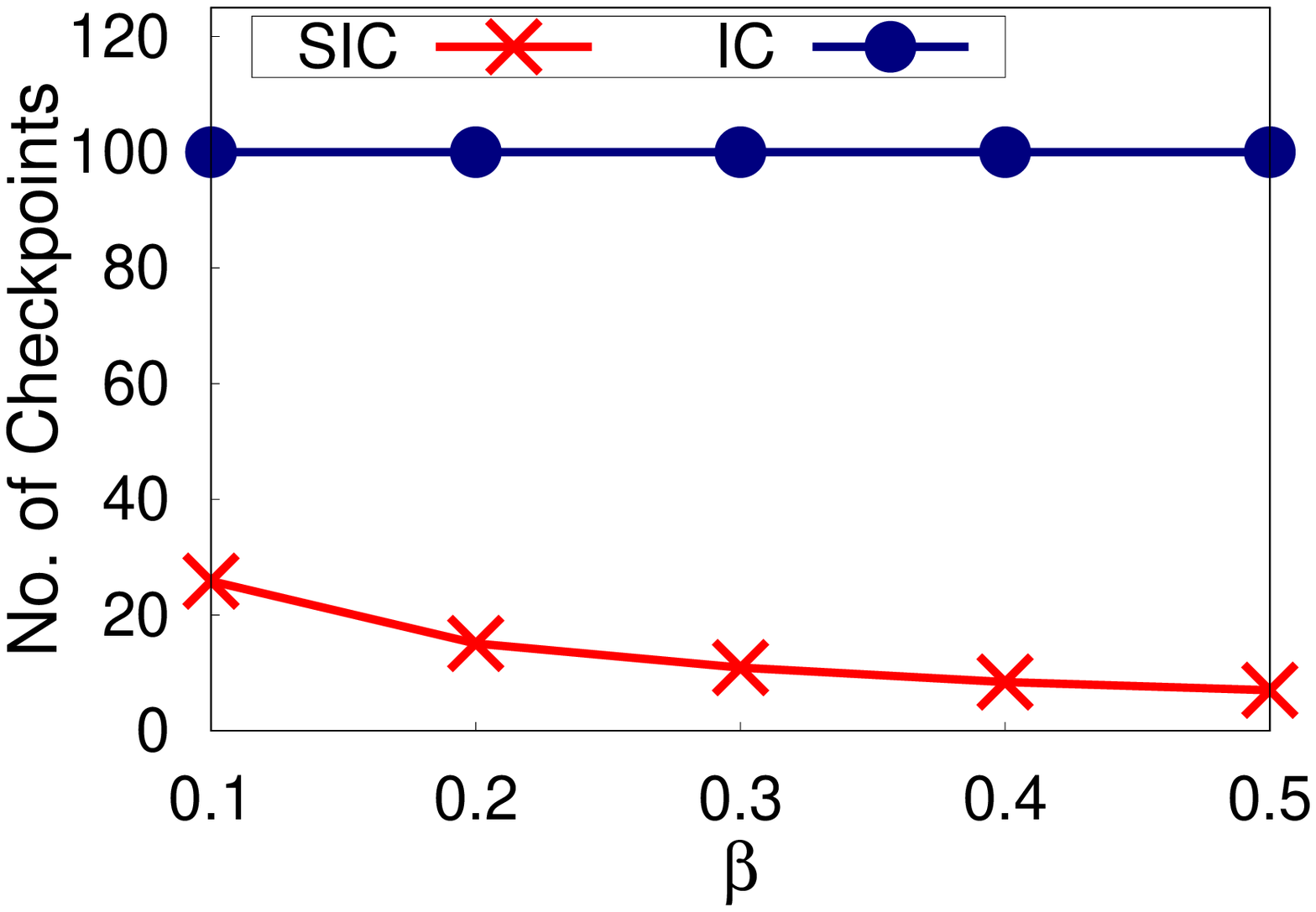}
        \label{fig:syn-n-e-checkpoints}
    }
    \caption{The number of checkpoints maintained by IC and SIC with varying $\beta$.}
    \label{fig:checkpointsBeta}
\end{figure*}
\begin{figure*}[htpb]
    \captionsetup[subfloat]{farskip=1pt,captionskip=1pt}
    \centering
    \subfloat[Reddit]{%
        \includegraphics[width=0.235\textwidth]{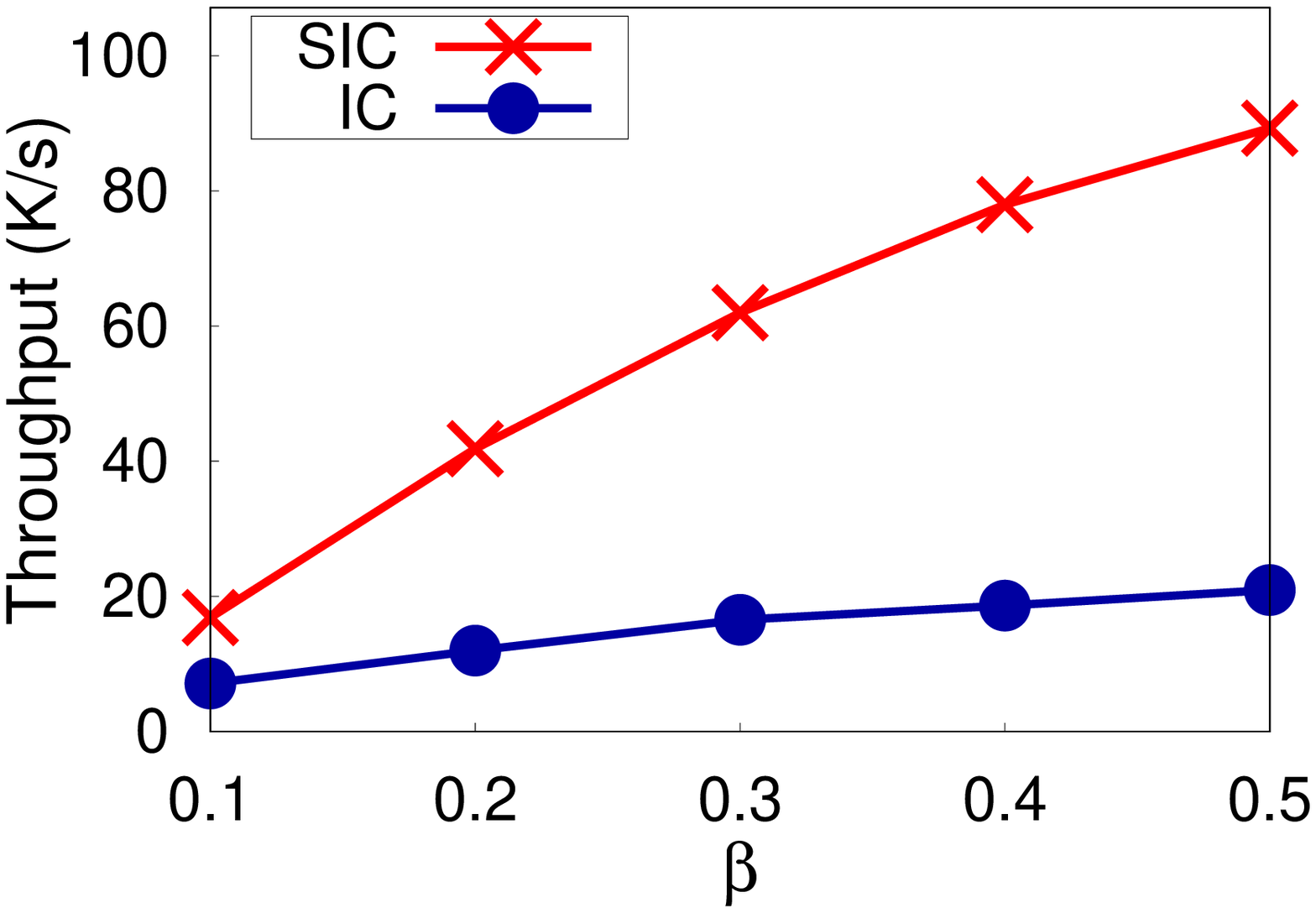}
        \label{fig:reddit-e-performance}
    }
    \subfloat[Twitter]{%
        \includegraphics[width=0.235\textwidth]{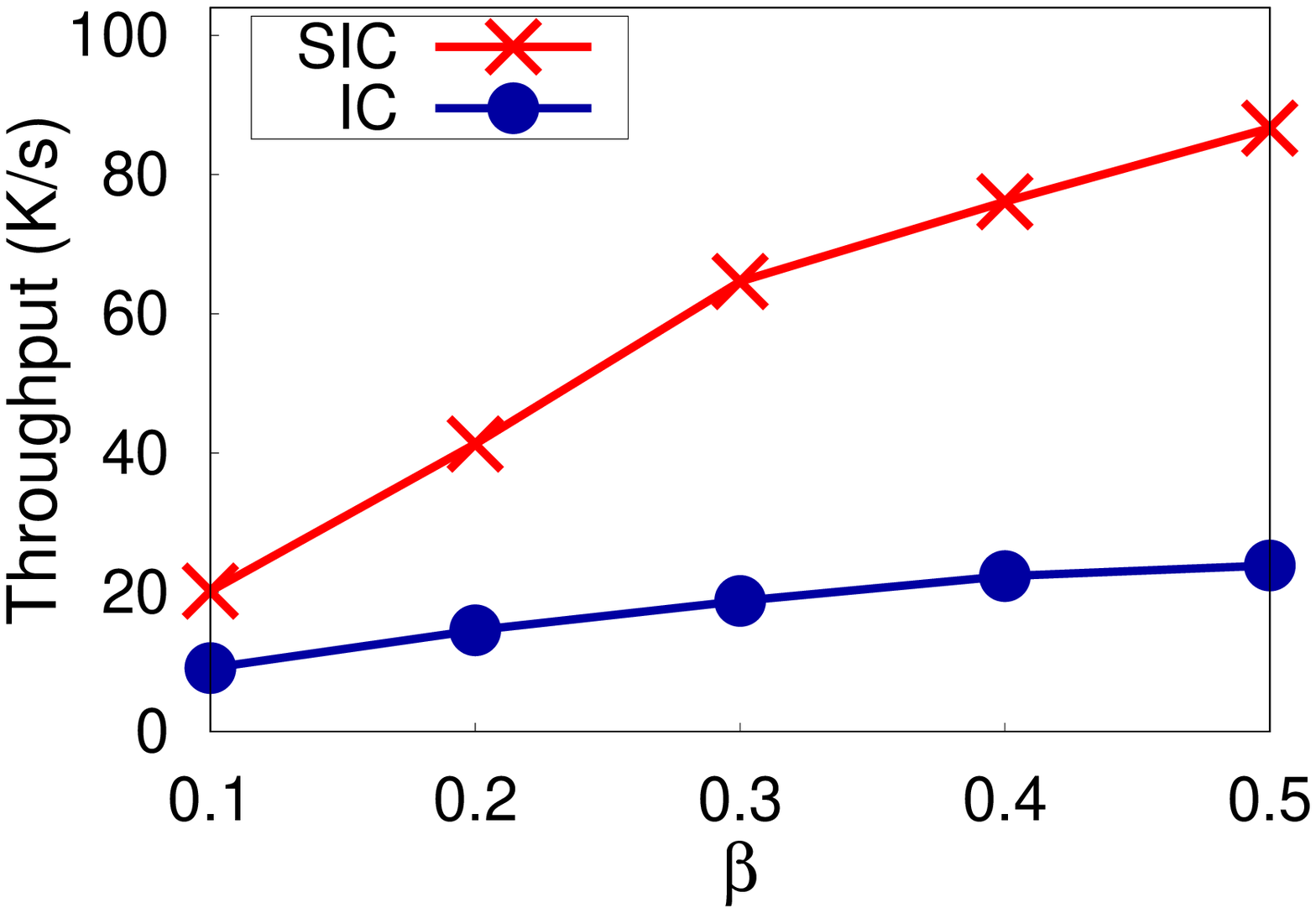}
        \label{fig:twitter-e-performance}
    }
    \subfloat[SYN-O]{%
        \includegraphics[width=0.235\textwidth]{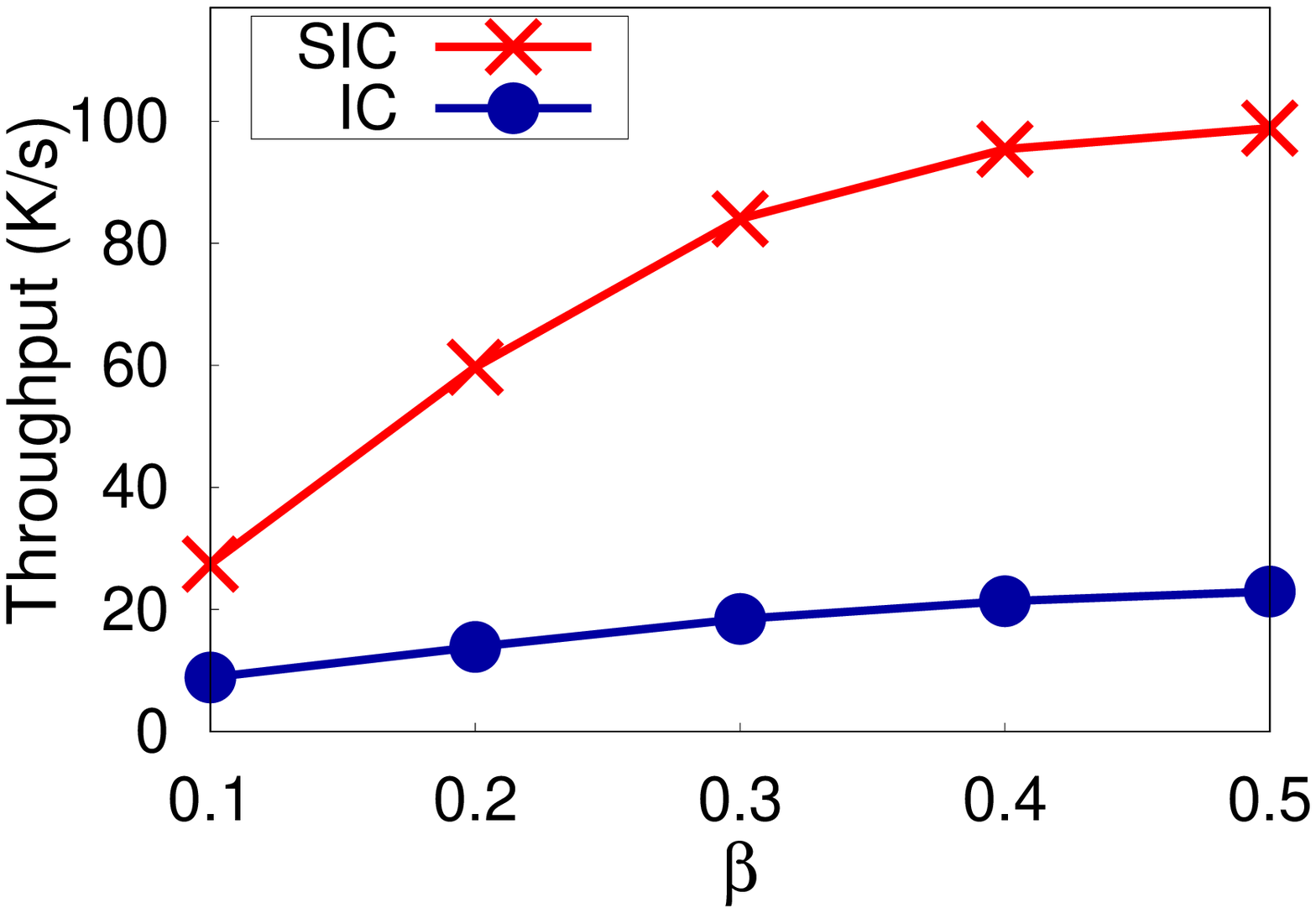}
        \label{fig:syn-o-e-performance}
    }
    \subfloat[SYN-N]{%
        \includegraphics[width=0.235\textwidth]{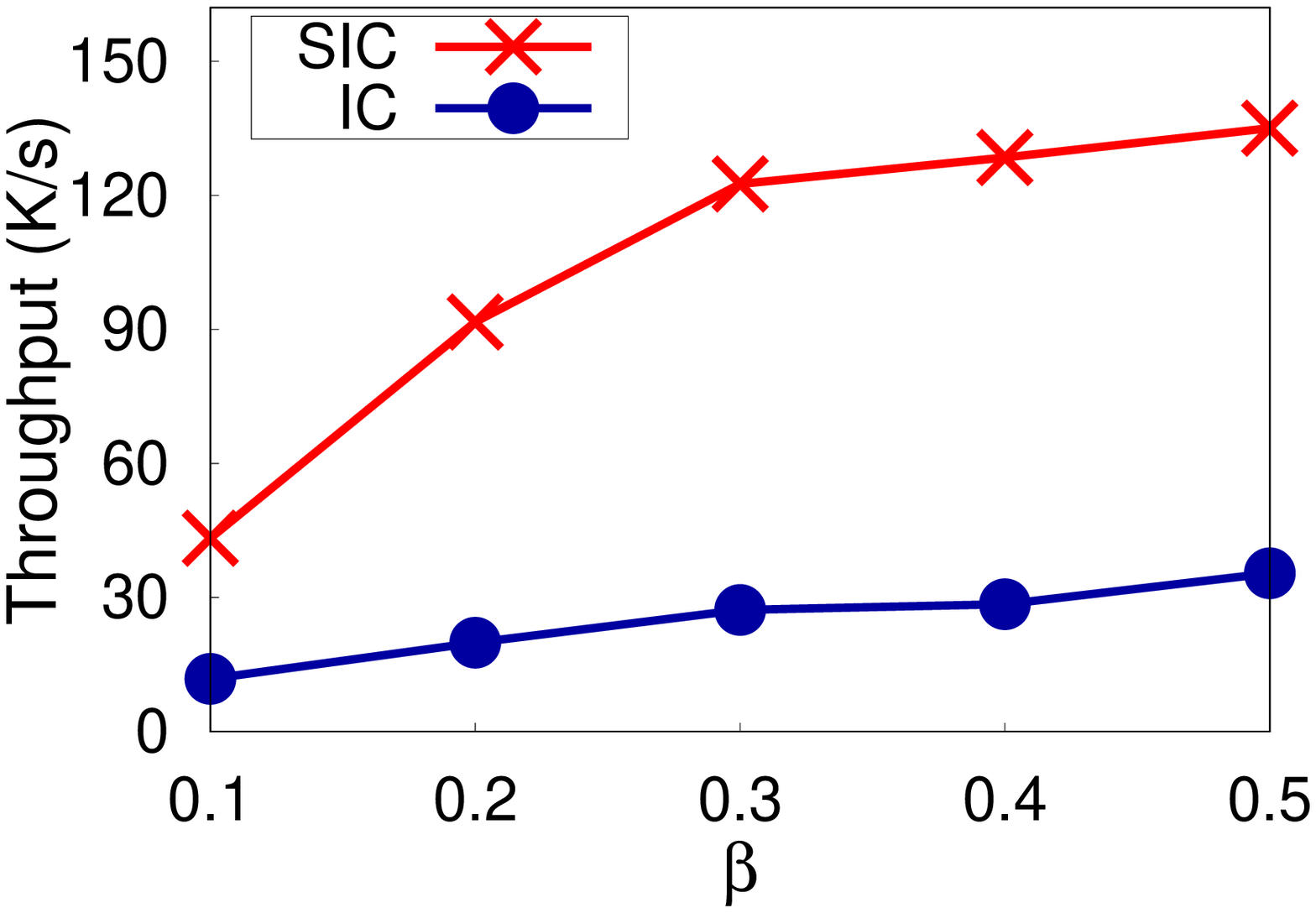}
        \label{fig:syn-n-e-performance}
    }
    \caption{Throughputs of IC and SIC with varying $\beta$.}
    \label{fig:throughputsBeta}
\end{figure*}
\begin{figure*}[htpb]
    \captionsetup[subfloat]{farskip=1pt,captionskip=1pt}
    \centering
    \subfloat[Reddit]{%
        \includegraphics[width=0.235\textwidth]{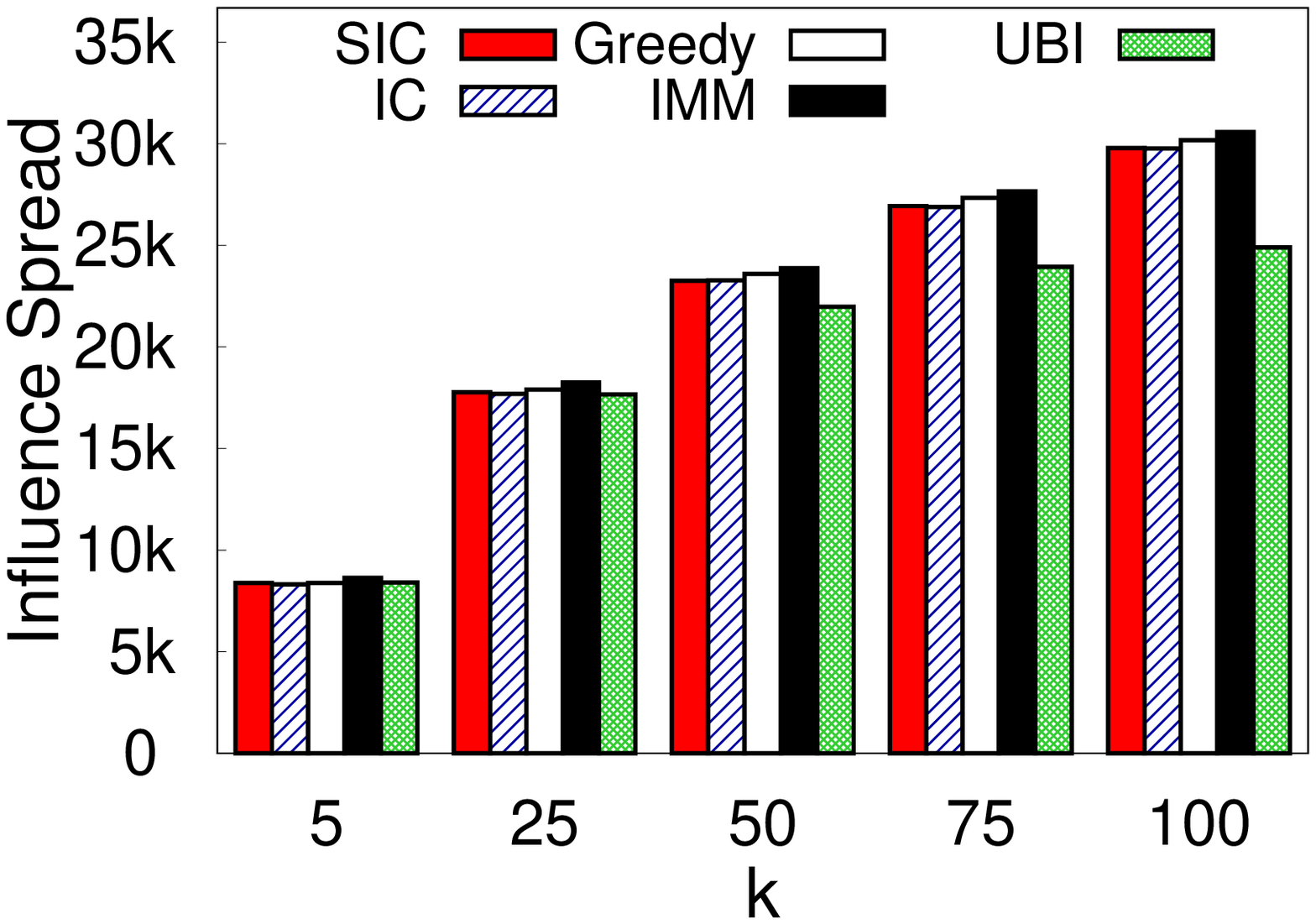}
        \label{fig:reddit-k-quality}
    }
    \subfloat[Twitter]{%
        \includegraphics[width=0.235\textwidth]{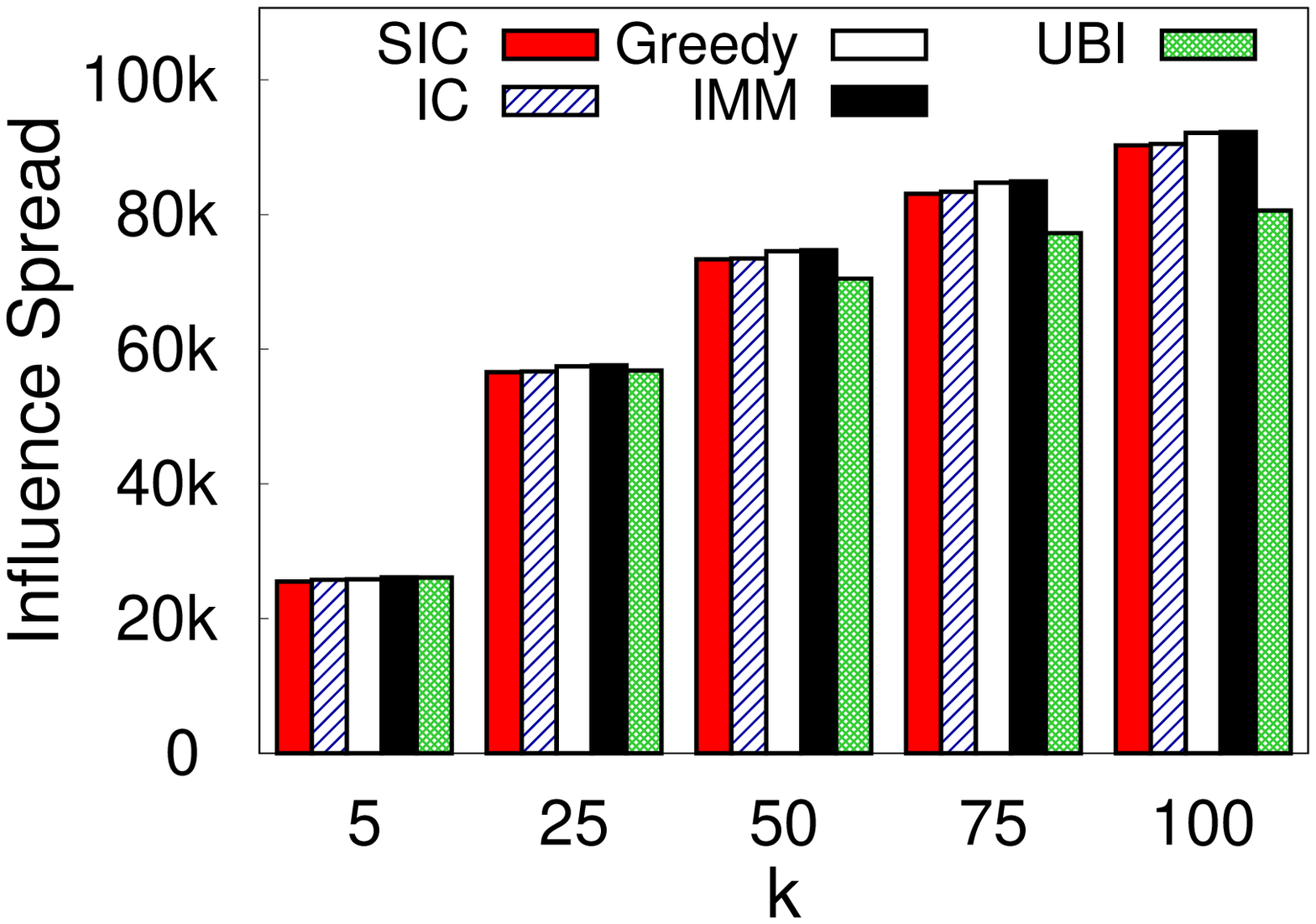}
        \label{fig:twitter-k-quality}
    }
    \subfloat[SYN-O]{%
        \includegraphics[width=0.235\textwidth]{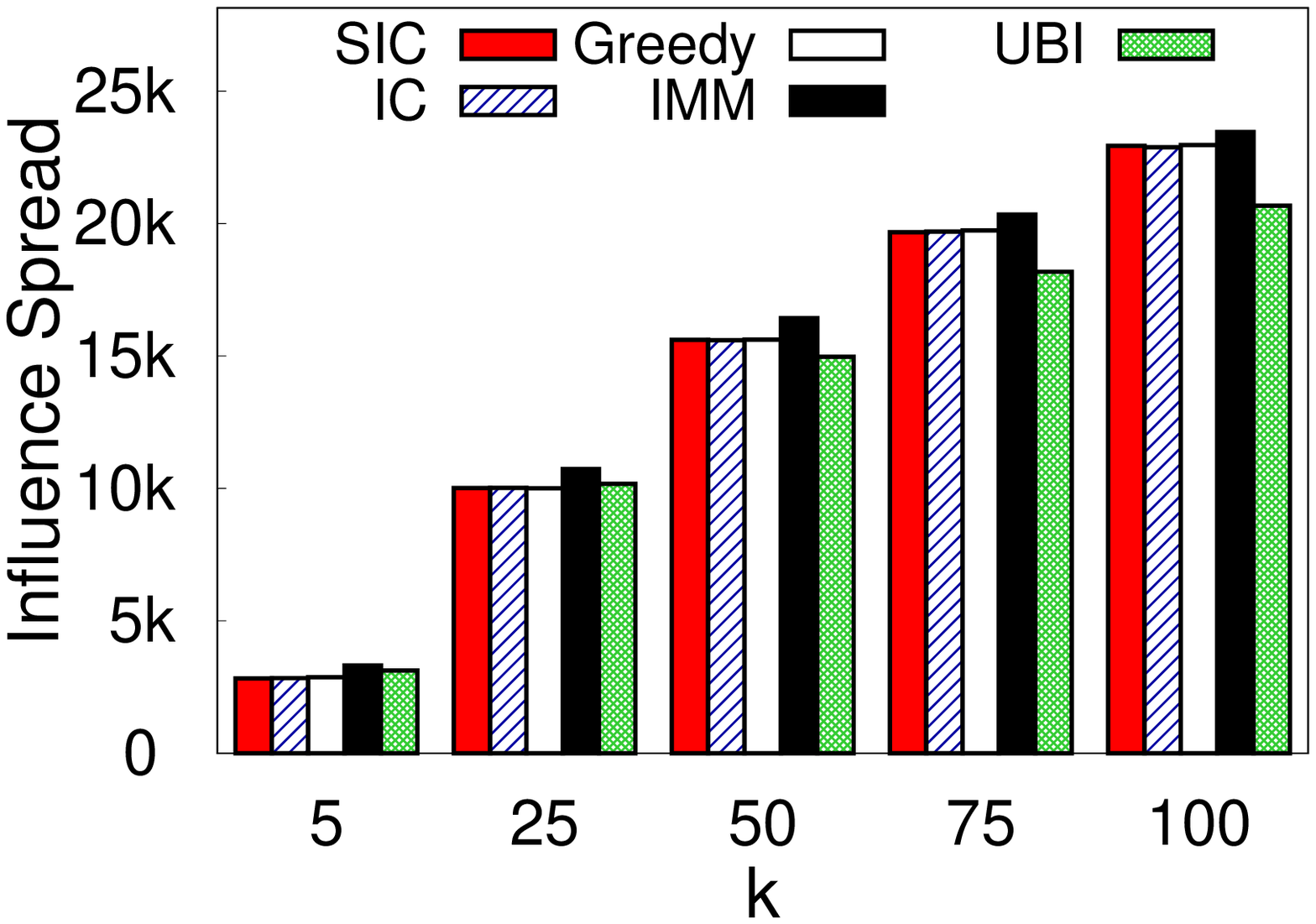}
        \label{fig:syn-o-k-quality}
    }
    \subfloat[SYN-N]{%
        \includegraphics[width=0.235\textwidth]{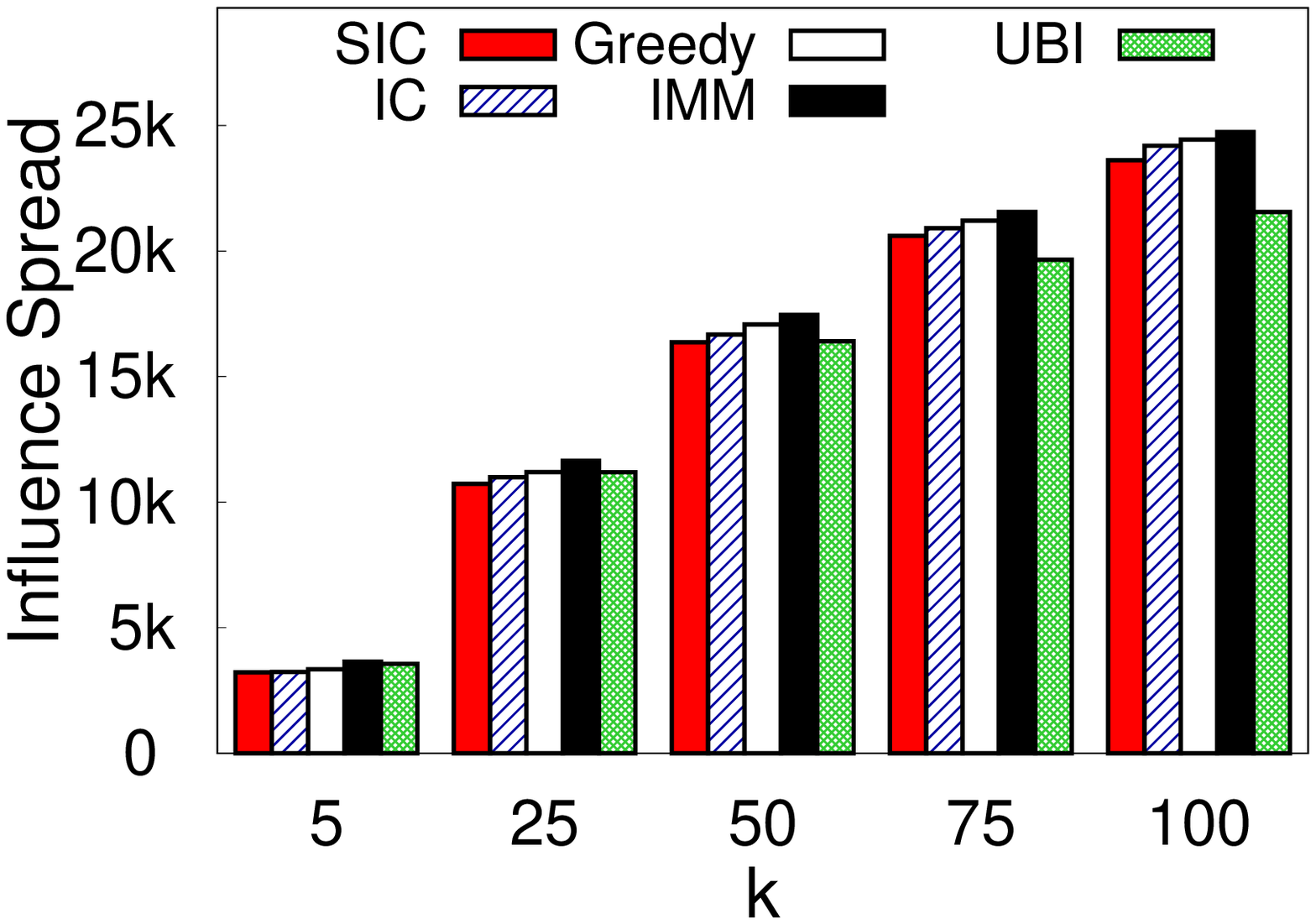}
        \label{fig:syn-n-k-quality}
    }
    \caption{Solution qualities of compared methods with varying $k$.}
    \label{fig:qualitiesK}
\end{figure*}
\begin{figure*}[htpb]
    \captionsetup[subfloat]{farskip=1pt,captionskip=1pt}
    \centering
    \subfloat[Reddit]{%
        \includegraphics[width=0.235\textwidth]{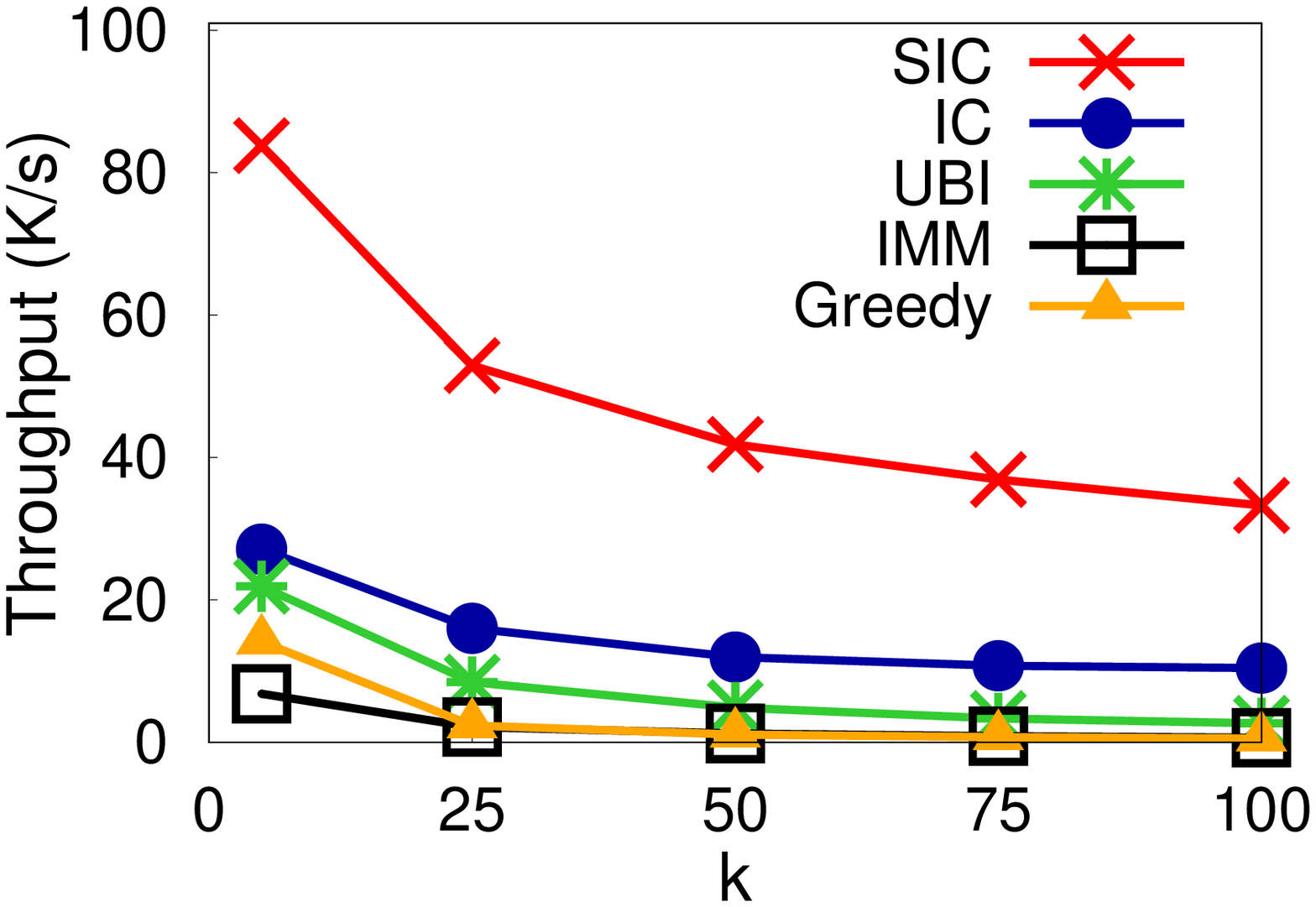}
        \label{fig:reddit-k-performance}
    }
    \subfloat[Twitter]{%
        \includegraphics[width=0.235\textwidth]{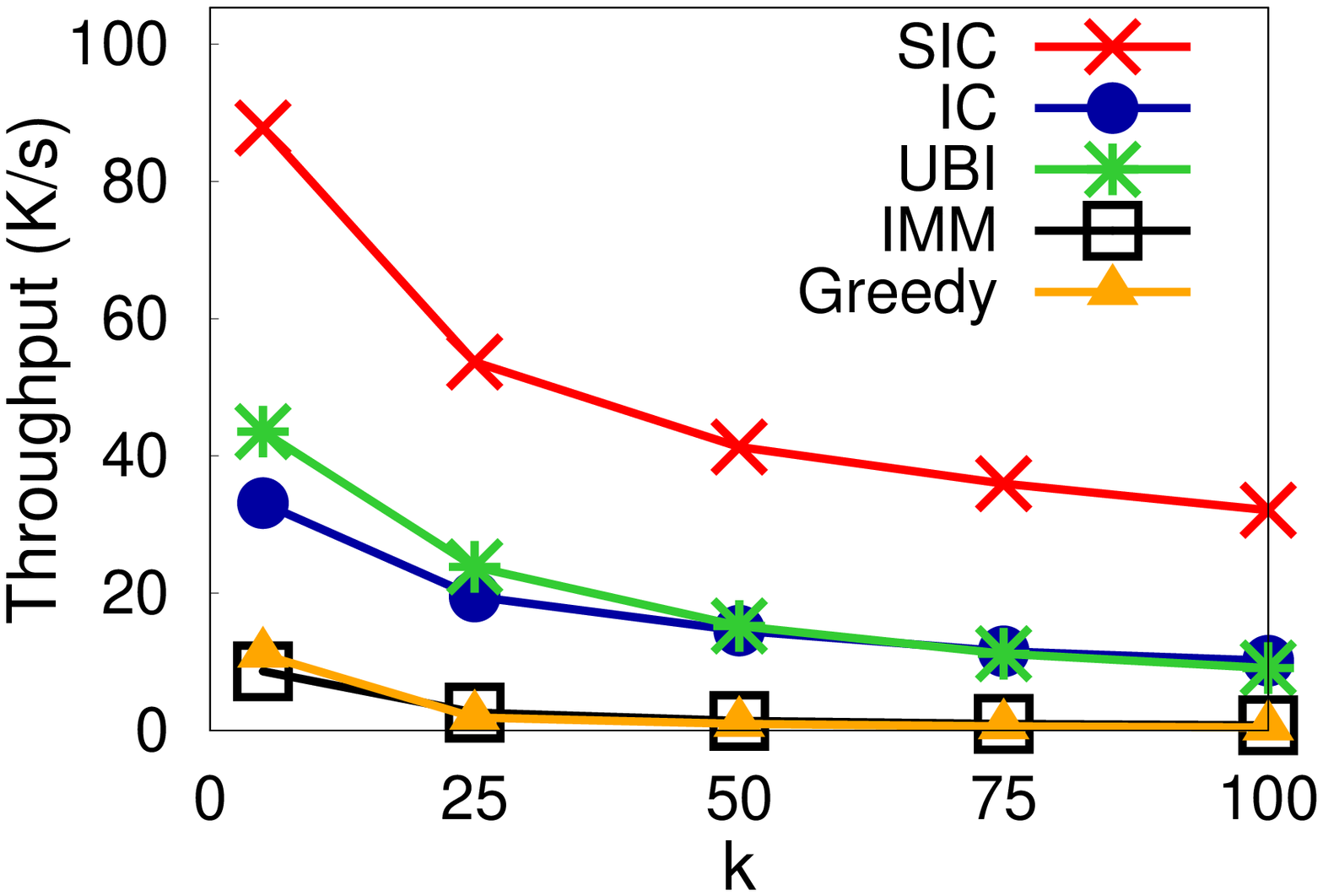}
        \label{fig:twitter-k-performance}
    }
    \subfloat[SYN-O]{%
        \includegraphics[width=0.235\textwidth]{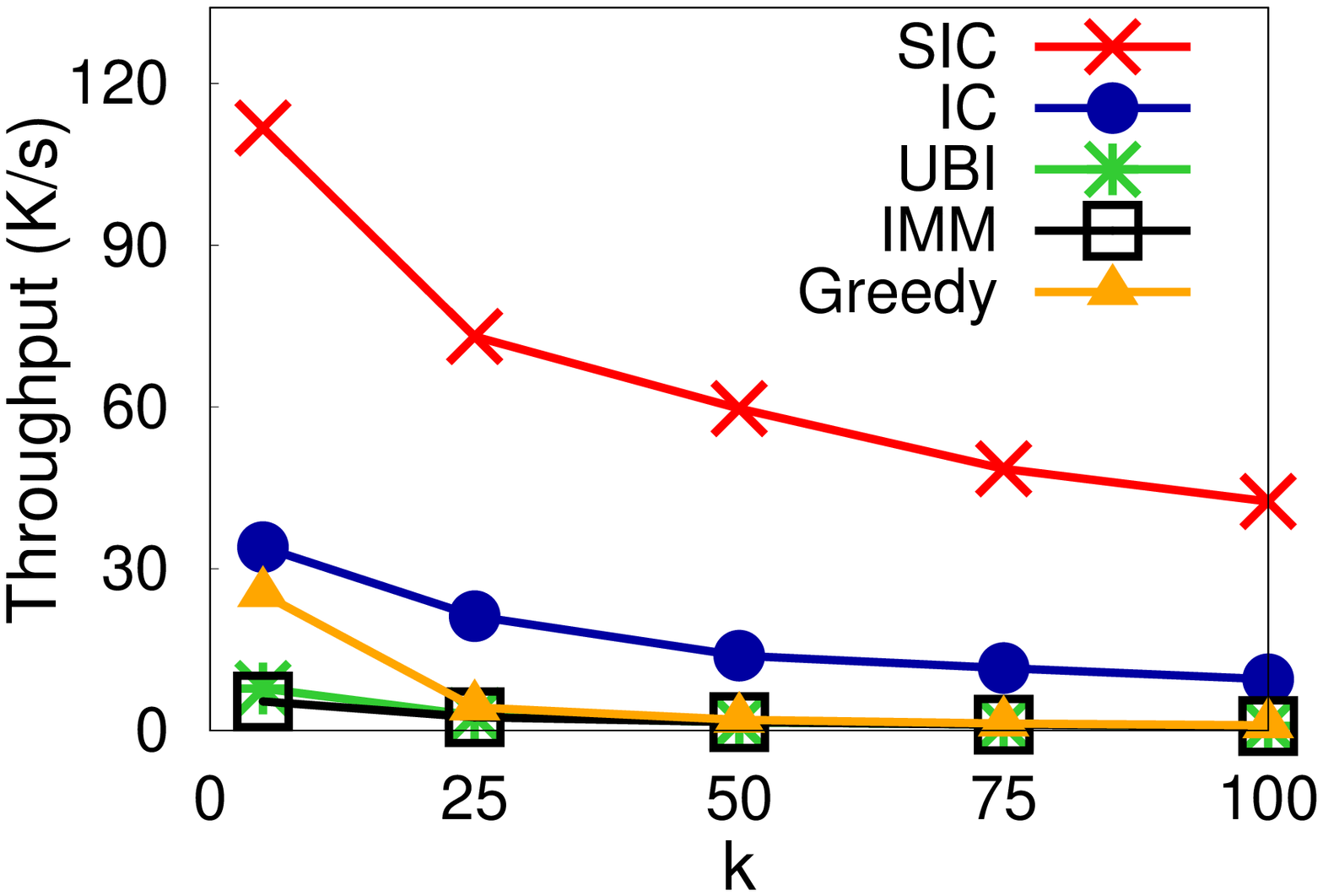}
        \label{fig:syn-o-k-performance}
    }
    \subfloat[SYN-N]{%
        \includegraphics[width=0.235\textwidth]{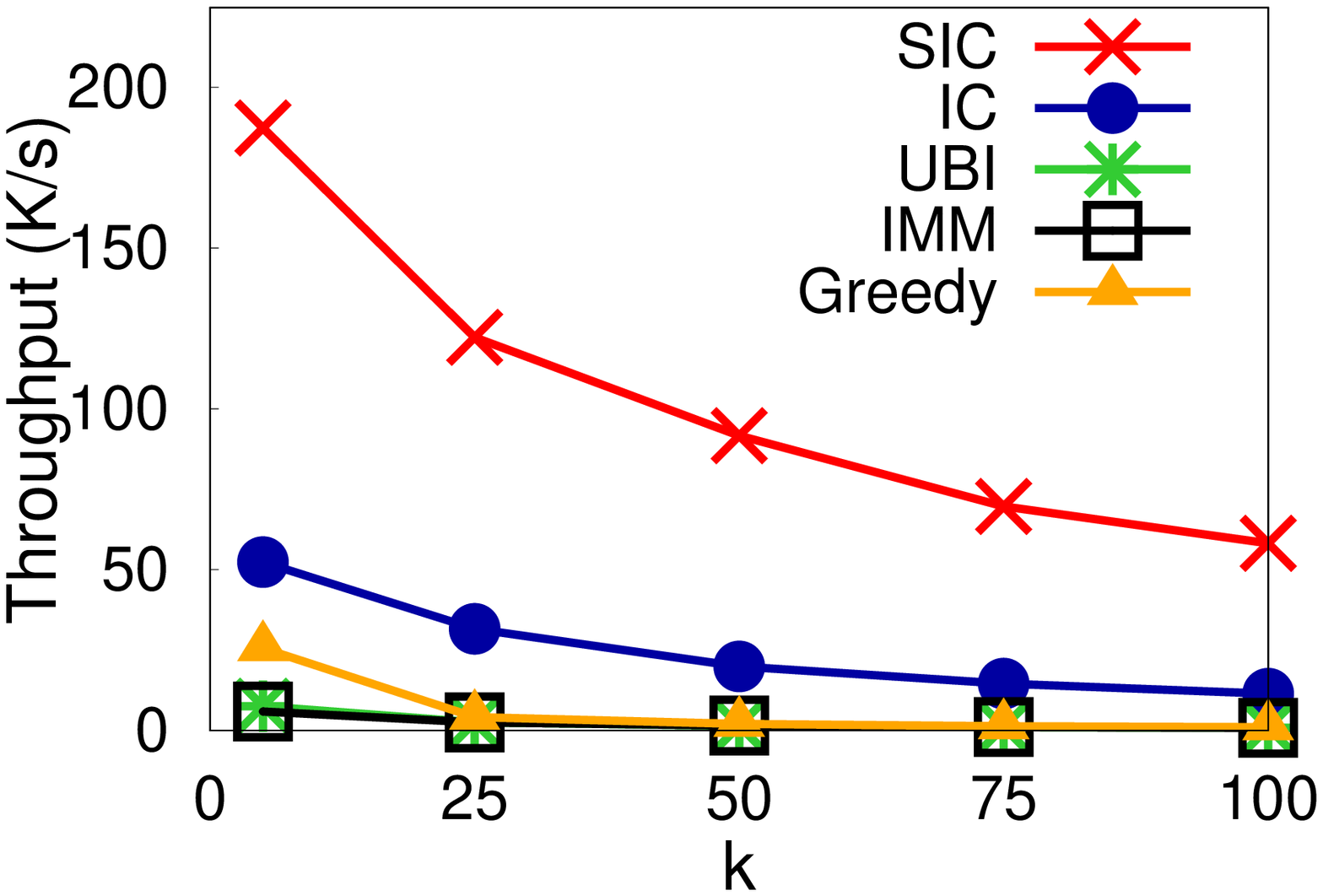}
        \label{fig:syn-n-k-performance}
    }
    \caption{Throughputs of compared methods with varying $k$.}
    \label{fig:throughputsK}
\end{figure*}
\begin{figure*}[htpb]
    \captionsetup[subfloat]{farskip=1pt,captionskip=1pt}
    \centering
    \subfloat[Reddit]{%
        \includegraphics[width=0.235\textwidth]{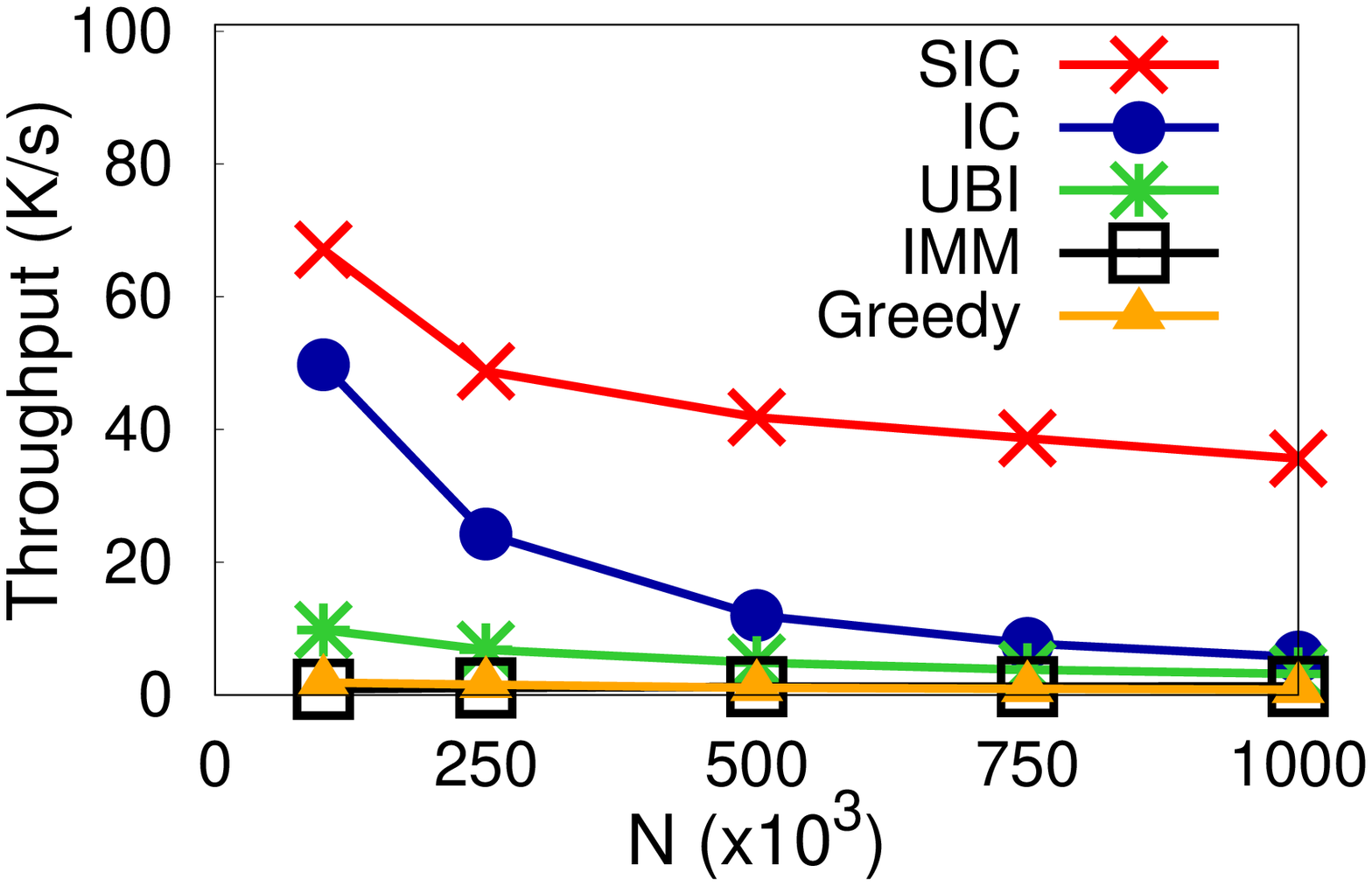}
        \label{fig:reddit-w-performance}
    }
    \subfloat[Twitter]{%
        \includegraphics[width=0.235\textwidth]{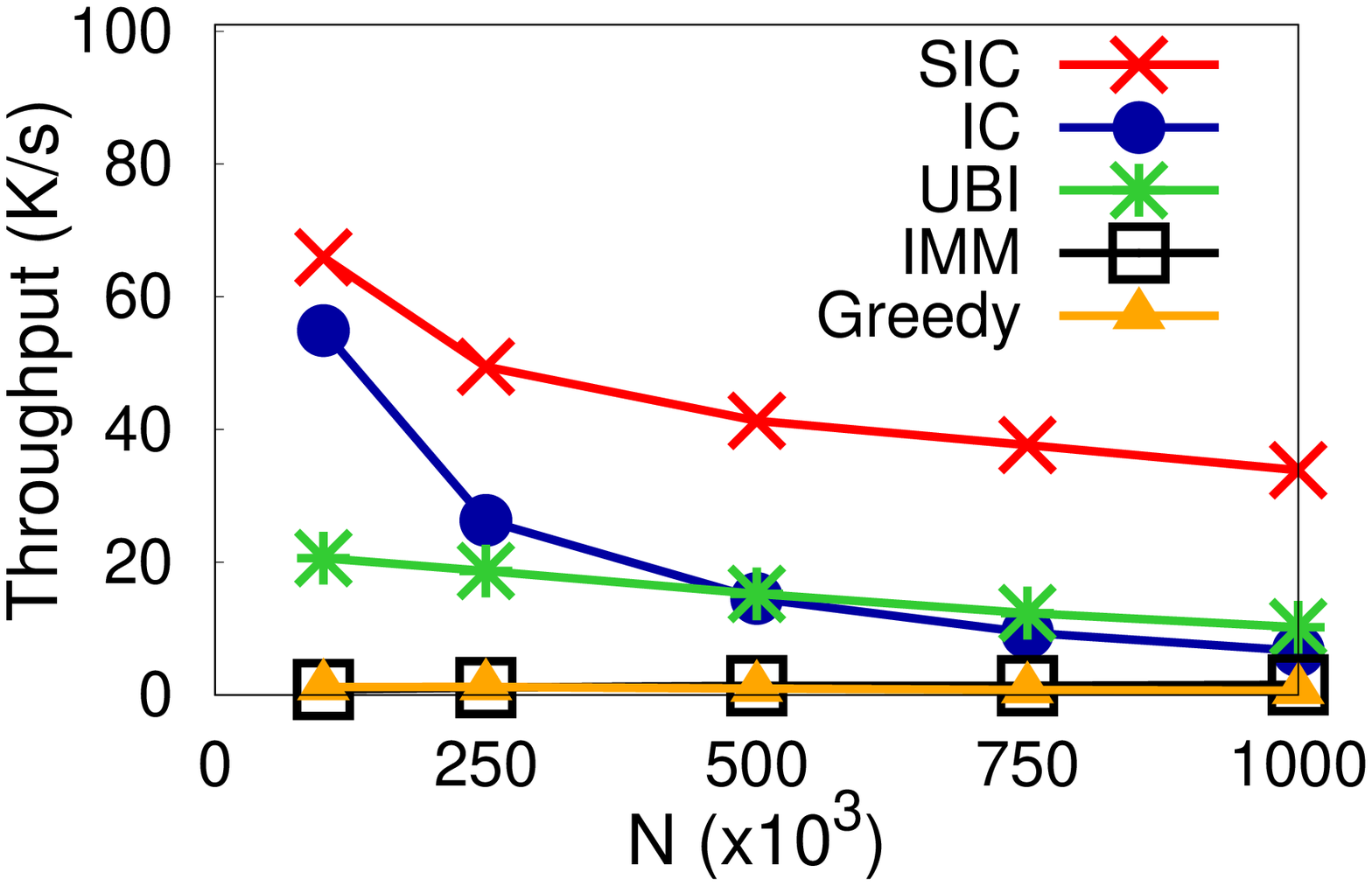}
        \label{fig:twitter-w-performance}
    }
    \subfloat[SYN-O]{%
        \includegraphics[width=0.235\textwidth]{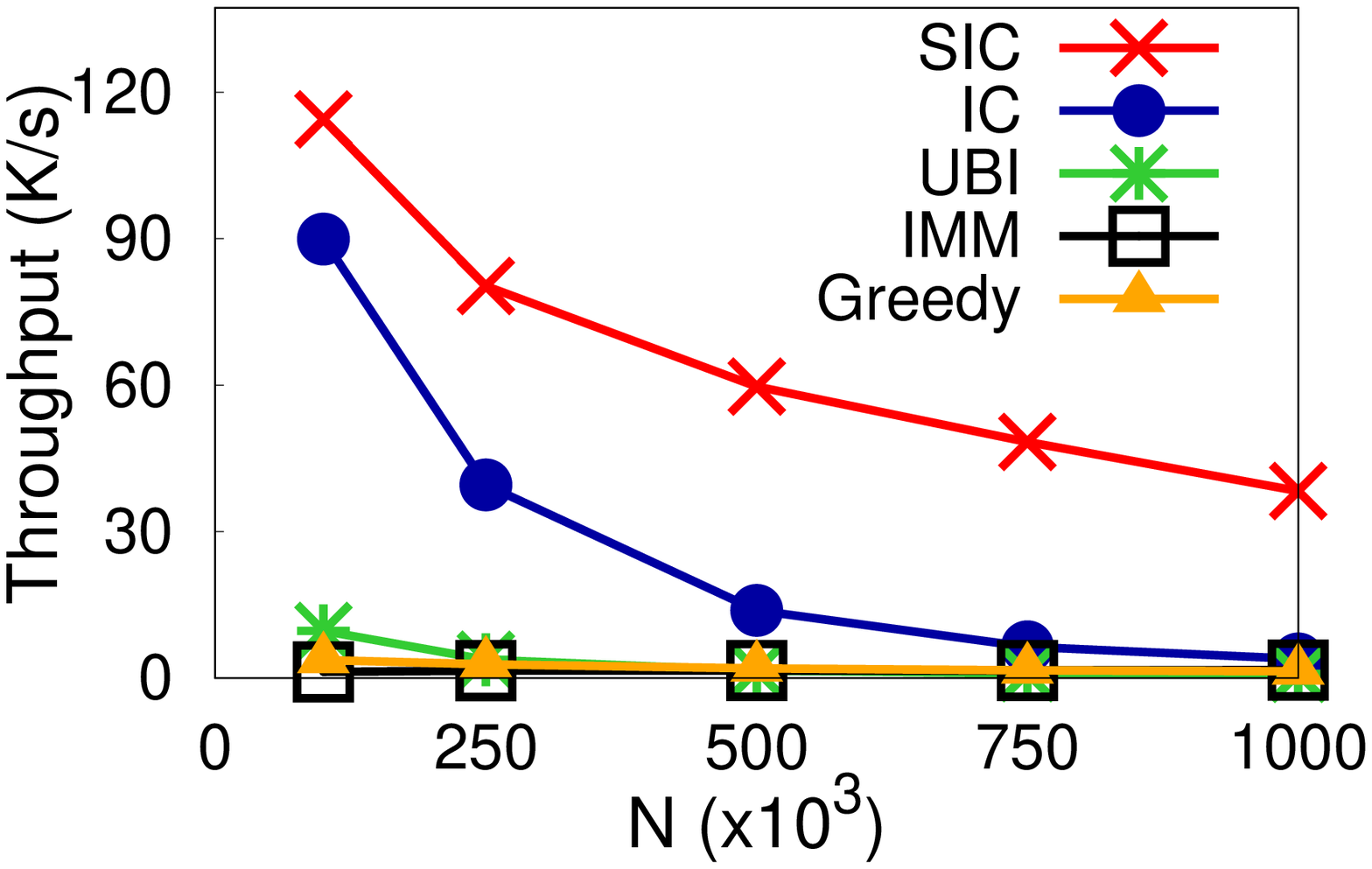}
        \label{fig:syn-o-w-performance}
    }
    \subfloat[SYN-N]{%
        \includegraphics[width=0.235\textwidth]{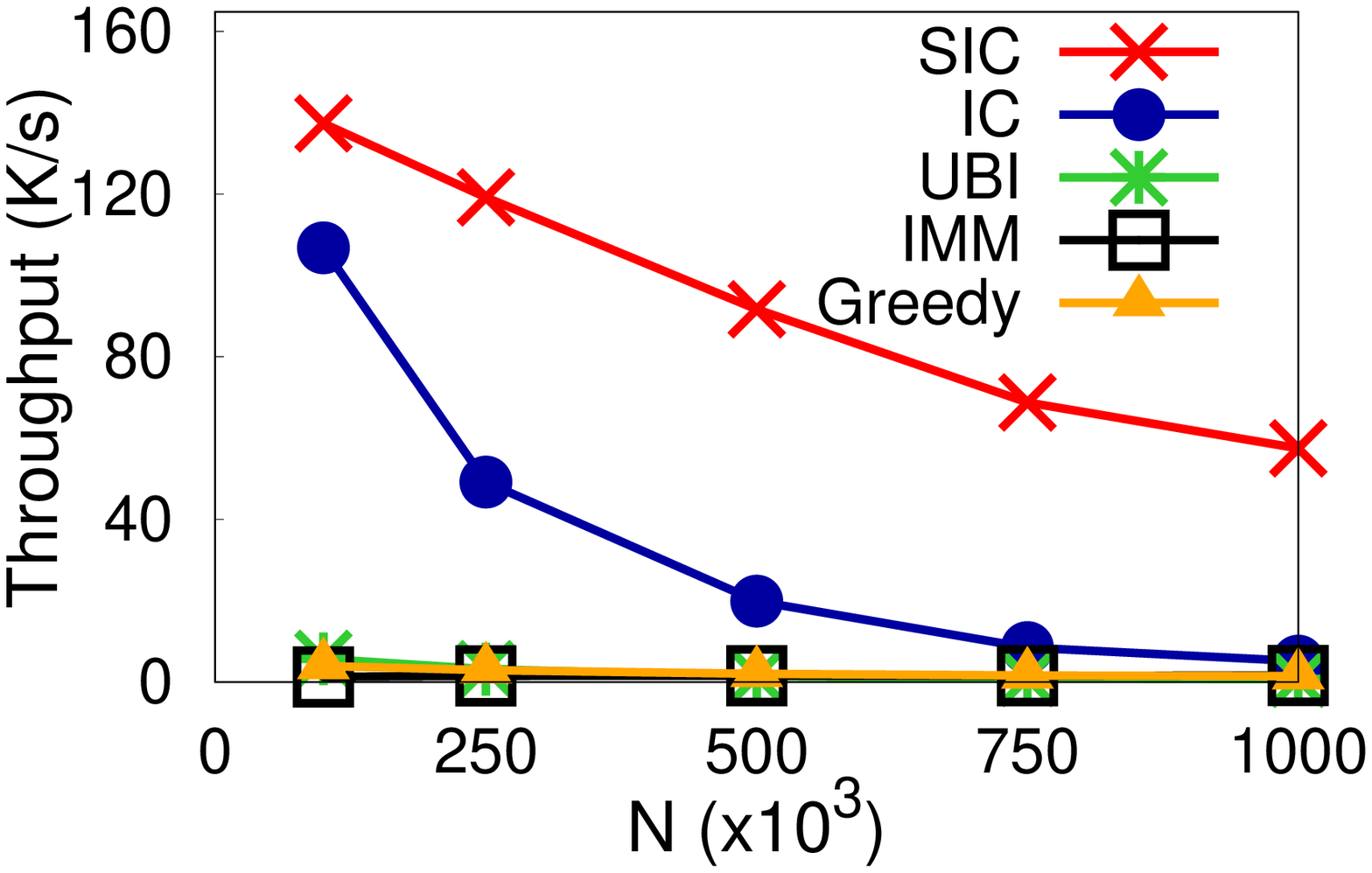}
        \label{fig:syn-n-w-performance}
    }
    \caption{Throughputs of compared approaches with varying $W$.}
    \label{fig:varyingW}
\end{figure*}
\begin{figure*}[htpb]
    \captionsetup[subfloat]{farskip=1pt,captionskip=1pt}
    \centering
    \subfloat[Reddit]{%
        \includegraphics[width=0.235\textwidth]{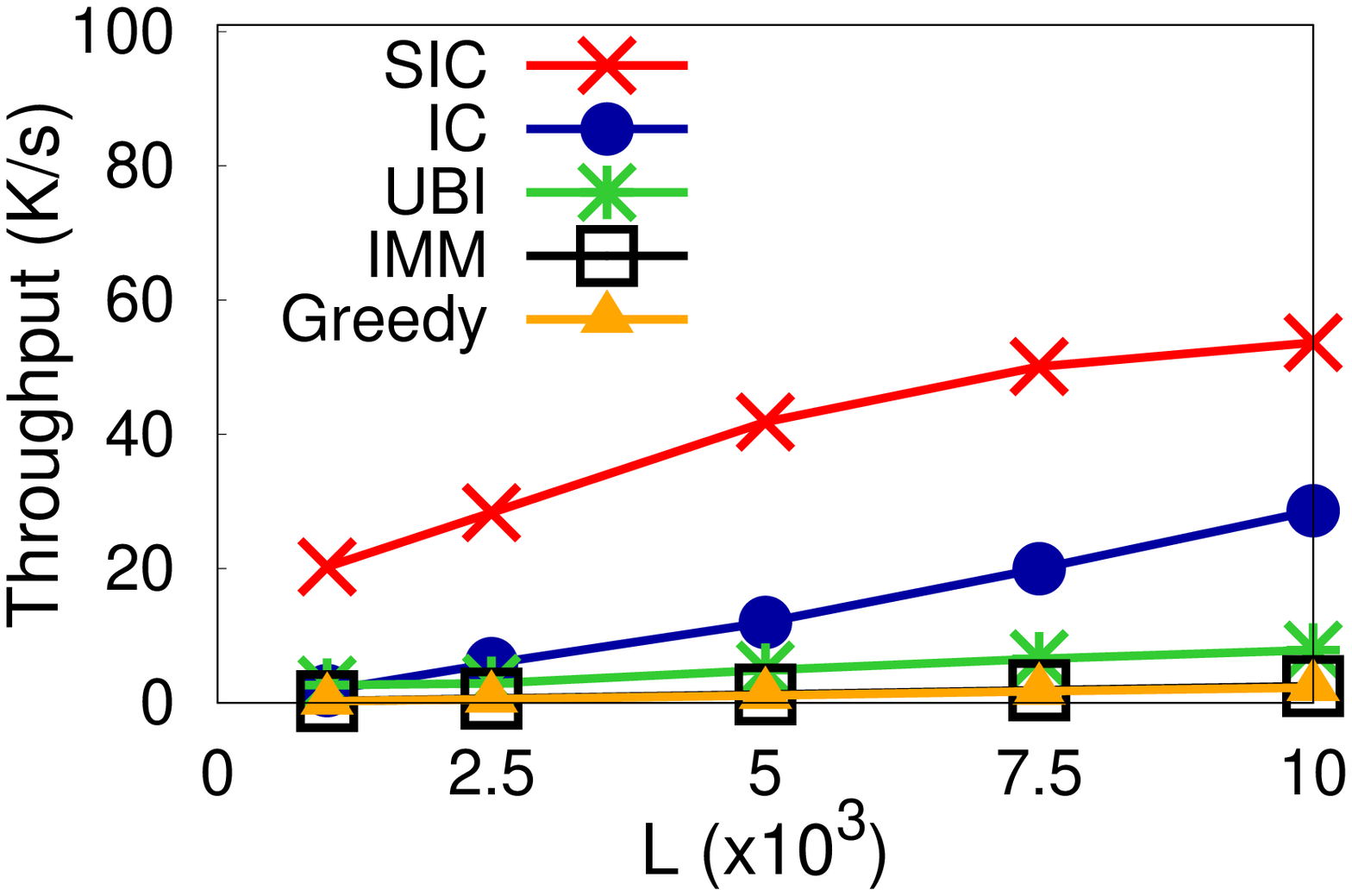}
        \label{fig:reddit-s-performance}
    }
    \subfloat[Twitter]{%
        \includegraphics[width=0.235\textwidth]{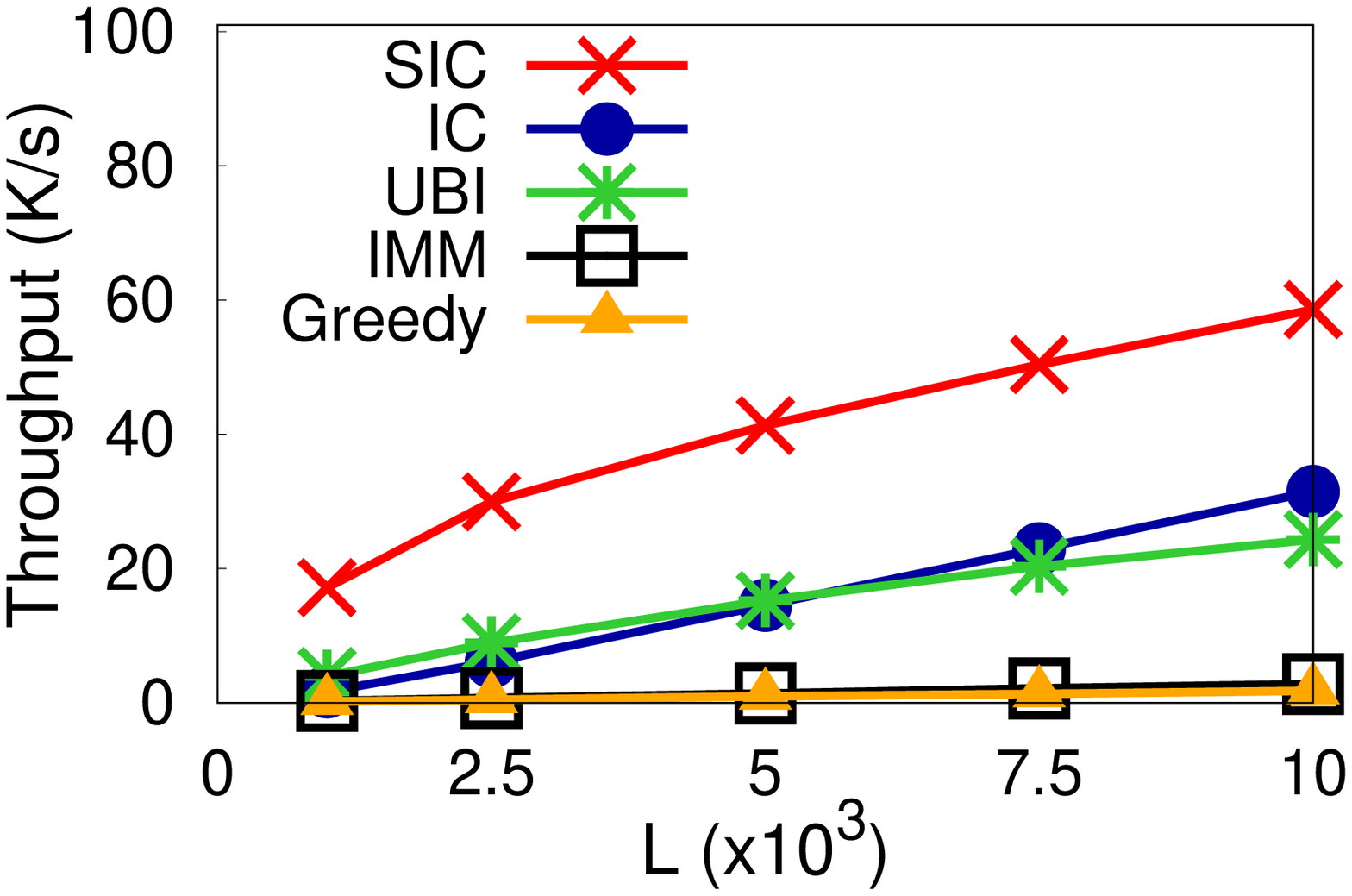}
        \label{fig:twitter-s-performance}
    }
    \subfloat[SYN-O]{%
        \includegraphics[width=0.235\textwidth]{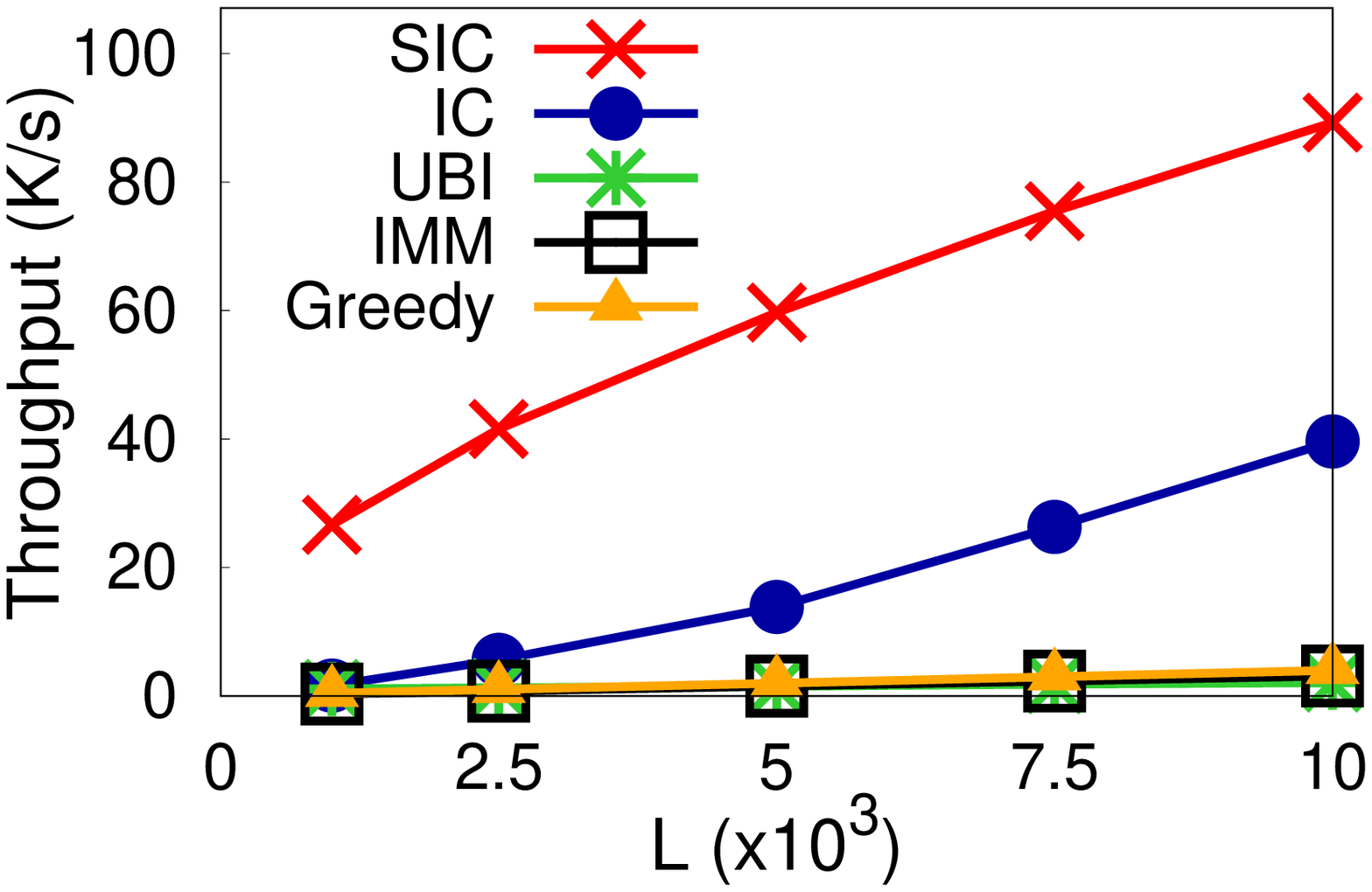}
        \label{fig:syn-o-s-performance}
    }
    \subfloat[SYN-N]{%
        \includegraphics[width=0.235\textwidth]{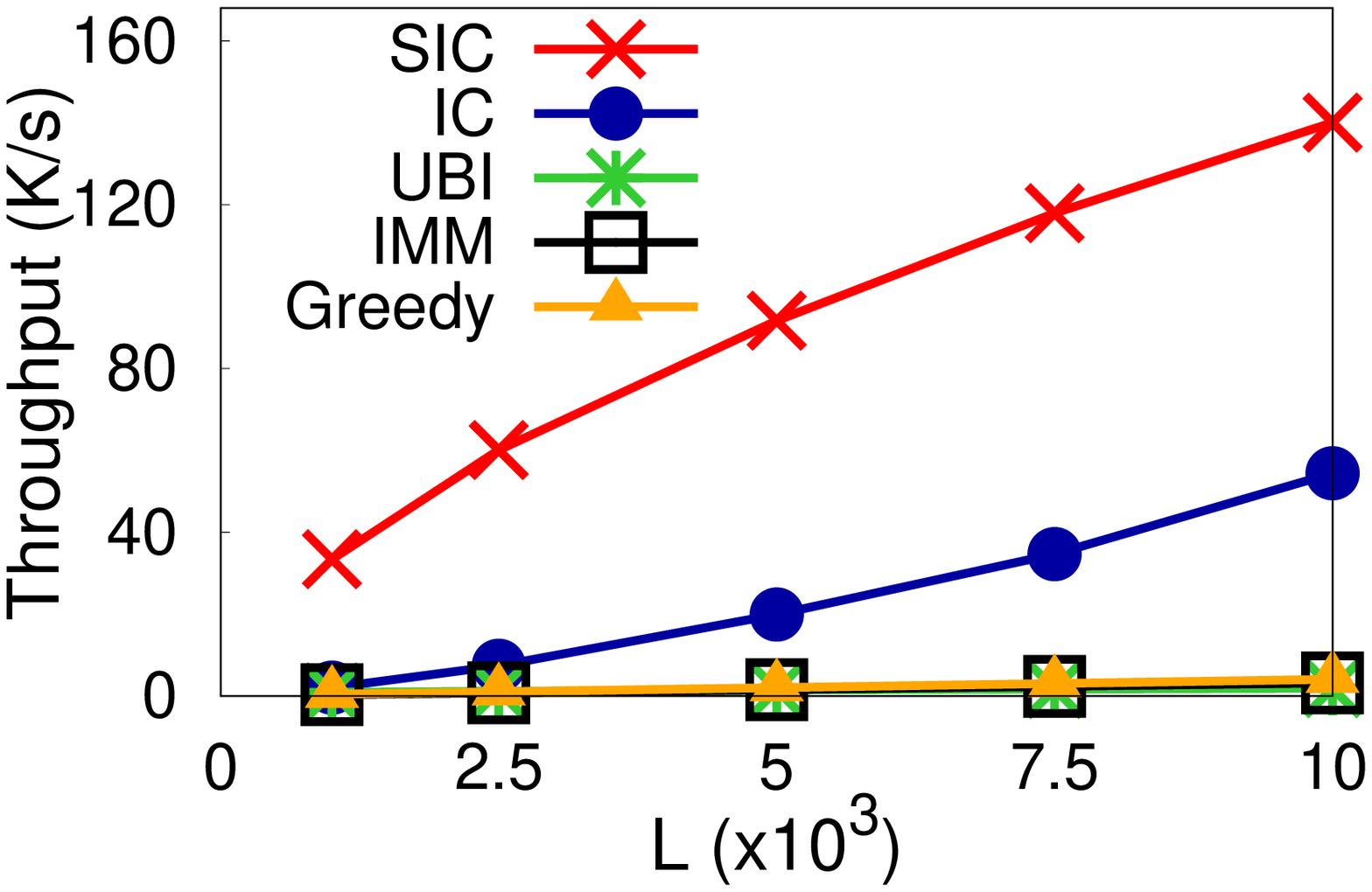}
        \label{fig:syn-n-s-performance}
    }
    \caption{Throughputs of compared approaches with varying $L$.}
    \label{fig:varyingL}
\end{figure*}

\noindent\textbf{Approaches}:
All approaches compared in the experiments are listed as follows:
\begin{itemize}[nosep,leftmargin=*]
    \item{
        \textbf{IMM}~\cite{Tang2015Influence}:
        To support our argument on the effectiveness, we use the state-of-the-art IM algorithm
        on static graphs as a baseline.
        At each time $t$, we construct an influence graph $G_t$ by treating users as vertices
        and the {\em influence} relationships between users wrt.~$W_t$ as directed edges.
        The edge probabilities between users are assigned by
        the weighted cascade (WC)~\cite{Kempe2003Maximizing} model.
        To extract the influential users at time $t$,
        we set the parameters of IMM to be $\varepsilon=0.5$, $l=1$~\cite{Tang2015Influence}
        and run the algorithm on the generated influence graph $G_t$.
    }
    \item{
        \textbf{UBI}~\cite{Chen2015Influential}:
        We use the state-of-the-art method for IM on dynamic graphs as another baseline.
        The generation of influence graphs is the same as \emph{IMM}.
        Then, a sequence of influence graphs $\{G_1,G_2,\ldots\}$ are fed to \emph{UBI}
        in a chronological order to track the influential users.
        We keep the same interchange threshold as used in~\cite{Chen2015Influential},
        i.e., $\gamma=0.01$.
    }
    \item{
        \textbf{Greedy}~\cite{Nemhauser1978Analysis}:
        We also implement the classic greedy algorithm in~\cite{Nemhauser1978Analysis}
        since it achieves the best theoretical approximation (i.e., $1-1/e$) of SIM queries.
        A detailed description of this algorithm is presented in Section~\ref{sec:ic}.
        Since the \emph{Greedy} algorithm does not store any intermediate result,
        it always recomputes the solution when being queried.
    }
    \item{
        \textbf{IC}:
        The IC framework proposed in Section~\ref{sec:ic}.
        We use \textsc{SieveStreaming}~\cite{Badanidiyuru2014Streaming} as the checkpoint oracle.
    }
    \item{
        \textbf{SIC}:
        The SIC framework proposed in Section~\ref{sec:sic}.
        We use the same checkpoint oracle as IC.
    }
\end{itemize}
\noindent\textbf{Quality Metric}:
We note that \emph{IMM} and \emph{UBI} work under the WC model whereas
\emph{Greedy}, \emph{IC} and \emph{SIC} are proposed to answer SIM queries
in Section~\ref{sec:problem}.
To verify the effectiveness of our proposed solutions,
we retrieve the seed users returned by all approaches for each window slide.
When a set of seed users is returned by each approach at time $t$,
we evaluate the {\em influence spread} of the users
under the WC model with 10,000 rounds of Monte-Carlo simulation
on the corresponding influence graph $G_t$.
Finally, we use the {\em average influence spread} of all windows for each approach 
as the quality metric.

\noindent\textbf{Performance Metric}:
We use \emph{throughput} as our performance metric. 
Specifically, whenever the window slides for $L$ actions,
we measure the elapsed CPU time of each approach and the throughput
is $L$ divided by the elapsed time.
We do not measure the query processing time because all approaches
maintain the seed users explicitly and the time to retrieve them
is negligible. 

\noindent\textbf{Parameters}:
The parameters examined in our experiments:
(1) $\beta$ is the parameter in \emph{IC} and \emph{SIC}
to achieve a trade-off between quality and efficiency.
(2) $k$ is the size of the seed set.
(3) $N$ is the window size. 
(4) $L$ is the number of actions for each window slide. 
(5) $|U|$ is the total number of users for synthetic datasets.
We vary $N$, $L$ and $|U|$ to test the scalability of the compared approaches.
The summary of parameters is listed in Table~\ref{tbl:variables}.
The default values of all parameters are in bold.

\begin{table}[t]
    \centering
    \caption{Parameters in experiments}
    \label{tbl:variables}
    \begin{tabular}{|c|l|}
        \hline 
        {\bf Parameter} & {\bf Values} \\ \hline 
        $k$ & 5, 25, \textbf{50}, 75, 100 \\ \hline 
        $\beta$ &  0.1, \textbf{0.2}, 0.3, 0.4, 0.5 \\ \hline 
        $N$ &  100K, 250K, \textbf{500K}, 750K, 1,000K \\ \hline 
        $L$ &  1K, 2.5K, \textbf{5K}, 7.5K, 10K \\ \hline
        $|U|$ &  1M, \textbf{2M}, 3M, 4M, 5M \\ \hline 
    \end{tabular}
\end{table}

\noindent\textbf{Experiment Settings}:
All experiments are conducted on a desktop machine running Ubuntu 14.04
with a quad core 3.4 GHz Intel i7-2600 processor and 16 GB memory.
All the approaches except \emph{IMM} are implemented in Java $8$.
The \emph{IMM} implementation available\footnote{\url{https://sourceforge.net/projects/im-imm/}}
is written in C++.

\subsection{Testing $\beta$ for IC and SIC}
\label{subsec:beta}
We first vary $\beta$ to test its effect on \emph{IC} and \emph{SIC}
in terms of the average influence value of SIM queries using the cardinality function, 
the number of maintained checkpoints and the throughput.
Note that we compare the seed qualities of \emph{IC} and \emph{SIC} with the baselines
in Section~\ref{subsec:varyingk} and only focus on their effectiveness of answering SIM queries here.

\noindent\textbf{Influence Value}:
The influence values of \emph{IC} and \emph{SIC} with varying $\beta$ are
presented in Figure~\ref{fig:reddit-e-quality}--\ref{fig:syn-n-e-quality}.
The influence values of \emph{IC} are slightly better than \emph{SIC} in most experiments.
This is because \emph{SIC} trades quality for efficiency by maintaining fewer checkpoints.
In spite of that, \emph{SIC} is able to obtain competitive values
with at most $5\%$ off from \emph{IC}.
In addition, we can see that both \emph{SIC} and \emph{IC}
achieve better influence values for a smaller $\beta$
and the influence values of \emph{SIC} degrade faster than \emph{IC} for a larger $\beta$
due to the deletion of checkpoints.
We note that in the SYN-N dataset,
the influence values of \emph{SIC} degrade more severely than other datasets for a larger $\beta$.
This is because the average reply distance is very short,
which leads to the frequent changes of the influential users.
Nevertheless, \emph{SIC} still returns solutions within the theoretical bound stated in
Section~\ref{subsec:theoretical}.

\noindent\textbf{Number of checkpoints}:
We examine the average number of checkpoints maintained by \emph{IC} and \emph{SIC}
for all sliding windows.
The results are presented in Figure~\ref{fig:reddit-e-checkpoints}--\ref{fig:syn-n-e-checkpoints}.
\emph{IC} maintains a constant number of checkpoints wrt.~$\beta$.
This is because the number of checkpoints maintained by \emph{IC} in each sliding window
exactly equals to $\lceil \frac{N}{L} \rceil$.
On the contrary, the number of checkpoints in \emph{SIC} is $O(\frac{\log N}{\beta})$
according to Theorem~\ref{thm:complexity} in Section~\ref{subsec:theoretical},
and is thus negatively correlated with $\beta$.
The trend for the number of checkpoints emphasizes
the superiority of \emph{SIC} in both space and time efficiencies.

\noindent\textbf{Throughput}:
The throughputs of \emph{IC} and \emph{SIC}
are presented in Figure~\ref{fig:reddit-e-performance}--\ref{fig:syn-n-e-performance}.
Both \emph{IC} and \emph{SIC} achieve better performance for a larger $\beta$.
There are two reasons behind such an observation.
First, both approaches employ \textsc{SieveStreaming} as the checkpoint oracle
where fewer candidate instances are kept within each checkpoint for a larger $\beta$,
which makes the update time shorter for each checkpoint.
Second, \emph{SIC} maintains fewer checkpoints as $\beta$ becomes larger,
which naturally leads to shorter processing time.
Thus, for a larger $\beta$, \emph{SIC} shows even more superiority
over \emph{IC} in all experiments
because fewer checkpoints are maintained by \emph{SIC}.

\subsection{Comparing Approaches with Varying $k$}
\label{subsec:varyingk}
We compare different approaches by using the quality and the performance metric defined
in Section~\ref{subsec:setup} for varying $k$.

\noindent\textbf{Quality}:
The results of solution qualities for different approaches are presented
in Figure~\ref{fig:reddit-k-quality}--\ref{fig:syn-n-k-quality}.
Compared with \emph{IMM}, \emph{Greedy}, \emph{IC} and \emph{SIC}
achieve less than $10\%$ quality losses.
The results have verified the effectiveness of SIM
as the seeds for SIM queries achieve nearly equivalent influence spreads
as the seeds retrieved by IMM under the WC model.
Moreover, \emph{SIC} shows competitive qualities though it
maintains fewer checkpoints than \emph{IC}.
In contrast,
the qualities of \emph{UBI} are close to \emph{IMM} when $k$ is small (i.e., $k \leq 25$).
But its qualities degrade dramatically when $k$ increases.
This is because \emph{UBI} relies on interchanging users to maintain the influential users
against the updates of the influence graph.
It interchanges a user into the maintained influential user set only when
a substantial gain is achieved for the estimated influence spread
(i.e., $1\%$ of the total influence spread prior to the interchange).
Thus, for a larger $k$, it becomes harder for a user to be interchanged
since the total influence spread of the maintained user set is larger.
This results in the delays of interchanges and causes larger errors.

\noindent\textbf{Throughput}:
The performances with varying $k$
are presented in Figure~\ref{fig:reddit-k-performance}--\ref{fig:syn-n-k-performance}.
The throughputs of all approaches are inversely proportional to $k$.
\emph{IC} and \emph{SIC} both employ \textsc{SieveStreaming} as the checkpoint oracle,
each checkpoint maintains a number of candidate instances
and each instance contains up to $k$ candidate users.
When $k$ gets larger, it is more expensive to evaluate the influence function for each checkpoint.
This explains why the performances of \emph{IC} and \emph{SIC} drop while $k$ becomes larger.
Compared with \emph{IC} and all baseline methods,
\emph{SIC} shows significant advantages in efficiency for all experiments.
Moreover, \emph{SIC} dominates \emph{Greedy} and \emph{IMM}
by achieving a speedup of up to 2 orders of magnitude across all datasets.
The throughput of \emph{UBI} is also far behind \emph{SIC} and \emph{IC}
on all datasets except \emph{Twitter}, where it achieves an equivalent throughput compared to \emph{IC}.
Nonetheless, \emph{UBI} is still more than 3 times slower than \emph{SIC}.

\subsection{Scalability}
In this section, we evaluate the scalability of compared methods
through measuring the throughputs
when varying the window size $N$, the length of each window slide $L$
and the total number of users $|U|$.
\begin{figure}[t]
    \centering
    \subfloat[SYN-O]{%
        \includegraphics[width=0.23\textwidth]{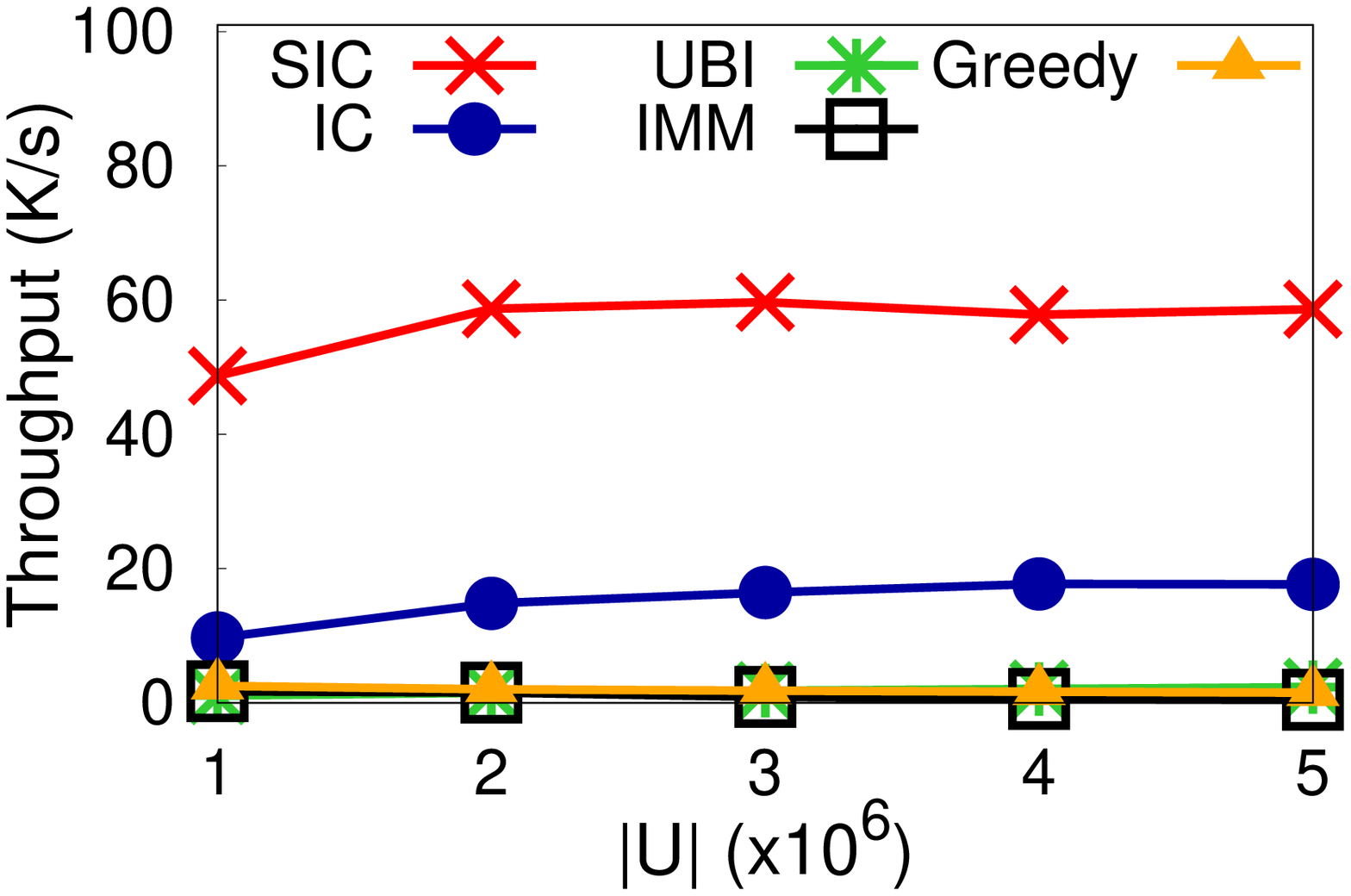}
        \label{fig:syn-o-user-performance}
    }
    \subfloat[SYN-N]{%
        \includegraphics[width=0.23\textwidth]{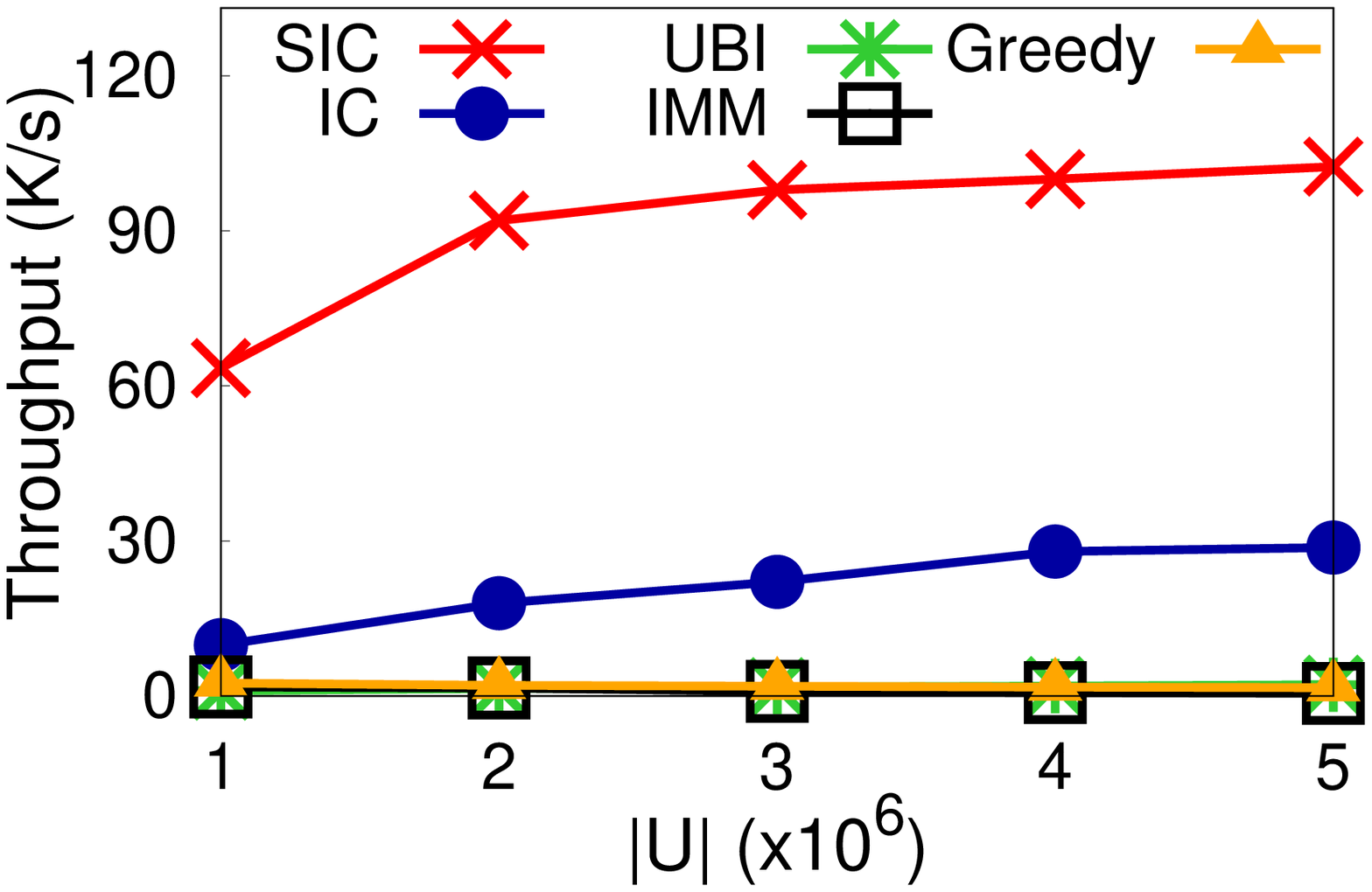}
        \label{fig:syn-n-user-performance}
    }
    \caption{Throughputs with varying $|U|$.}
    \label{fig:user}
\end{figure}

\noindent\textbf{Varying $N$}:
The performances for varying $N$ are presented
in Figure~\ref{fig:reddit-w-performance}--\ref{fig:syn-n-w-performance}.
Although the throughputs of all approaches decrease with increasing $N$,
\emph{SIC} shows better scalability as it only maintains $O(\log{N})$ checkpoints
when $\beta$ is fixed.
We observe a smaller performance gap between \emph{IC} and \emph{SIC}
in all datasets when $N$ is small (i.e., $N=100,000$).
This is because the number of checkpoints maintained by
\emph{IC} is very close to \emph{SIC} (fewer than $8$)
and the benefits of sparse checkpoints become less significant.
Nonetheless, when $N$ increases, \emph{SIC} regains its superiority.
Moreover, \emph{SIC} achieves speedups of up to
40x, 100x and 70x compared to 
\emph{Greedy}, \emph{IMM} and \emph{UBI} respectively.

\noindent\textbf{Varying $L$}: 
We show the performances for varying $L$ in
Figure~\ref{fig:reddit-s-performance}--\ref{fig:syn-n-s-performance}. 
As $L$ increases, the throughputs of \emph{IC} and \emph{SIC} increase. This 
is because larger $L$ results in a smaller number of checkpoints for both methods. 
\emph{IC} exhibits a linear performance improvement wrt.~larger $L$ since it maintains $\lceil \frac{N}{L} \rceil$ checkpoints.
As \emph{SIC} deletes some checkpoints created by \emph{IC},
it continues to be superior to \emph{IC} in terms of performance,
which demonstrates its scalability in handling multiple window slides.
Like the results for varying $N$, 
\emph{SIC} dominates \emph{Greedy}, \emph{IMM} and \emph{UBI} in terms of the throughput. 

\noindent\textbf{Varying $|U|$}:
Finally, we show the performances for varying $|U|$ on two synthetic datasets
in Figure~\ref{fig:syn-o-user-performance}--\ref{fig:syn-n-user-performance}.
We observe that the throughputs of \emph{SIC}, \emph{IC} and \emph{UBI} increase
as $|U|$ becomes larger.
Fixing the window size $N$, the influence graphs become more sparse for a larger $|U|$.
All these three approaches show better performance on more sparse graphs.
However, \emph{Greedy} and \emph{IMM} need more processing time to run as $|U|$ increases
since these complexities are directly related to $|U|$.
Finally, \emph{SIC} still shows superior performances in all experiments with various $|U|$ settings.

\section{Conclusion} 

In this paper, we proposed a novel \emph{Stream Influence Maximization} (SIM) query to retrieve
$k$ influential users who collectively maximize the influence value over a social action stream.
Then, we presented a novel framework \emph{Influential Checkpoints} (IC)
and its improved version \emph{Sparse Influential Checkpoints} (SIC)
to efficiently support the continuous SIM queries over high-speed social streams.
Theoretically, SIC maintained $O(\frac{\log N}{\beta})$ checkpoints
to obtain an $\frac{\epsilon(1-\beta)}{2}$-approximate solution for SIM queries.
Empirically, our experiments showed that SIC achieves
up to $2$ orders of magnitude speedups over the state-of-the-art static and dynamic IM approaches
with less than $10\%$ losses in seed quality.
In particular, SIC demonstrated a peak processing rate of more than $150$K actions per second,
which is adequate for real-world social streams.
In the future, we plan to extend our proposed frameworks
to support a broader class of IM problems,
e.g., competitive IM~\cite{Bharathi2007Competitive,Li2015GetReal,Lu2015CCC}.

\section*{Acknowledgment}

We thank Mr. Ohsaka Naoto for providing his implementation of algorithms in~\cite{Ohsaka2016Dynamic}.

\bibliographystyle{abbrv}
\bibliography{sec9_references}

\appendix
\section{Discussion}
\label{sec:discussion}

In this section, we discuss how to extend our proposed frameworks
to different types of IM problems in a streaming setting.
Here, we take topic-aware IM~\cite{Barbieri2012Topic,Chen2015OTI},
location-aware IM~\cite{Li2014ELI,Wang2016Distance},
and conformity-aware IM~\cite{Li2015Conformity} as examples.

\noindent{\bf Topic-Aware IM}:
First, we consider processing topic-aware SIM queries in a dynamic social stream.
Given a set of topics $T$, we assume that a topic oracle
can provide a subset of topics $T_t \subseteq T$ for each arrival action $a_t$,
where $T_t$ contains all topics that are related to $a_t$.
Then, we consider each SIM query $q$ only concerns a subset of topics $T_q \subseteq T$.
To maintain the influential users for $q$,
we can invoke IC or SIC but only process a sub-stream of actions $\{a_t|T_t \cap T_q \neq \emptyset\}$.

\noindent{\bf Location-Aware IM}:
Then, we consider the location-aware SIM queries in a dynamic social stream.
For a location-aware SIM query $q$ with a spatial region $R$,
we aim to find users who have the largest influences in $R$.
It can be processed in a similar way with topic-aware SIM queries.
We assume each action $a_t$ is associated with a position $p_t$,
which means that $a_t$ happens at $p_t$.
To maintain the influential users for $q$,
we can also invoke IC or SIC but only process a sub-stream of actions $\{a_t|p_t \in R\}$.

\noindent{\bf Conformity-Aware IM}:
In conformity-aware IM, the influence and conformity scores of all users are calculated offline.
To incorporate these scores into SIM,
we can design the influence function as follows:
given any user set $S$, the influence function is set to $\sum_{u' \in I(S)}w_S(u')$
where $I(S)$ is the set of users influenced by $S$ in our model
and $w_{S}(u')$ is the conformity-aware influence score from $S$ to a user $u'$.
Specifically, $w_S(u')$ is evaluated as $w_S(u')=1-\prod_{u \in S \wedge u' \in I(\{u\})}(1-\Phi(u)\Omega(u'))$
with $\Phi(\cdot)$ and $\Omega(\cdot)$ denoting the influence and conformity scores calculated by
the conformity-aware model offline. 
As conformity-aware scores naturally evolve slowly,
it is reasonable to treat them as constants for a long time and recompute them periodically.
As the adapted influence function retains the monotone and submodular properties,
our proposed frameworks remain effective and efficient when considering conformity-aware IM.

From the above examples, we see different types of IM problems with monotone submodular objective functions
can be adapted to our frameworks.
However, it is still an open problem to use our frameworks to support IM problems having non-monotone or
non-submodular objective functions (e.g. opinion-aware IM) or considering the dependency of different queries
(e.g., competitive IM).

\end{document}